\documentclass[a4paper, 11pt]{article}
\usepackage[pdftex, a4paper]{geometry}
\usepackage{amsfonts}
\usepackage{dsfont}
\usepackage{mathrsfs}
\usepackage{amssymb}
\usepackage{epsfig}
\usepackage{amsmath}
\usepackage{appendix}
\usepackage{float}
\usepackage{color}
\usepackage{epstopdf}
\usepackage[english]{babel}

\newtheorem{theorem}{Theorem}
\newtheorem{definition}{Definition}

\newtheorem{lemma}{Lemma}
\newtheorem{remark}{Remark}
\newtheorem{proposition}{Proposition}
\newtheorem{assumption}{Assumption}
\newtheorem{example}{Example}
\begin{document}

\title{\bf  Persistent Flows and Non-Reciprocal Interactions in  Deterministic Networks}

\author{Weiguo Xia,   Guodong Shi, Ziyang Meng, Ming Cao, and Karl Henrik Johansson\thanks{
W. Xia is with the School of Control Science and Engineering,
Dalian University of Technology, China ({\tt\small wgxiaseu@dlut.edu.cn}).} \thanks{G. Shi is with the Research School of Engineering, The Australian National University, Australia ({\tt\small guodong.shi@anu.edu.au}).} \thanks{Z. Meng is with the State Key Laboratory of Precision Measurement Technology and Instruments, Department of Precision Instrument, Tsinghua University, China ({\tt\small ziyangmeng@mail.tsinghua.edu.cn}).} \thanks{M. Cao is with the Faculty of Mathematics and Natural Sciences, ENTEG, University of Groningen, the Netherlands ({\tt\small m.cao@rug.nl}).} \thanks{K. H. Johansson is with ACCESS Linnaeus Centre, School of Electrical Engineering, Royal Institute of Technology, Sweden ({\tt\small kallej@kth.se}).}
}

\date{}
\maketitle

\begin{abstract}
This paper studies  deterministic consensus networks with discrete-time dynamics  under persistent flows and non-reciprocal agent  interactions. An arc describing the interaction strength between two agents is said to be persistent if its weight function has an infinite $l_1$ norm. We discuss two balance conditions on the interactions between agents which generalize the arc-balance and cut-balance conditions in the literature respectively. The proposed conditions require that such a balance should be satisfied over each time window of a fixed length instead of at each time instant. We prove that in both cases global consensus is reached if and only if the persistent graph, which consists of all the persistent arcs, contains a directed spanning tree. The convergence rates of the system to consensus are also provided in terms of the interactions between agents having taken place. The  results are obtained under a weak condition without assuming the existence of a positive lower bound of all the nonzero weights of arcs and are compared with the existing results. Illustrative examples are provided to show the critical importance of the nontrivial lower boundedness of the self-confidence of the agents.
\end{abstract}

\section{Introduction}

In distributed coordination of multi-agent systems, a great deal of attention has been paid to consensus-seeking systems. The study of this type of systems is motivated by opinion forming in social networks \cite{HeKr02,BlHeTs09}, flocking behaviors in animal groups \cite{ViCzJaCoSh95,Ol06}, data fusion in engineered systems \cite{BoGhPrSh06} and so on. Ample results on the convergence and convergence rate of the consensus system have been reported. Typical conditions involve the connectivity of the network topology and the interaction strengths between agents for both continuous-time \cite{OlMu04,ReBe05,MaGi13,HeTs13,ShJo13c,BoMa15} and discrete-time systems \cite{JaLiMo03,ReBe05,Mo05,CaMoAn08a,NeOlOzTs09,ShJo13c}.

In the literature, several types of balance conditions on the interaction weights are considered, among which the cut-balance condition \cite{HeTs13,MaGi13} and the arc-balance condition \cite{ShJo13c} are typical ones. The cut-balance condition requires that at each time instant, if a group of agents in the network influences the remaining ones then it is also influenced by the remaining ones bounded by a constant proportional amount. This type of conditions characterizes a reciprocal interaction relationship among  the agents, which covers the symmetric interaction and type-symmetric interaction as special cases \cite{HeTs13}. It was proved in \cite{HeTs13} that under the cut-balance condition, the state of the consensus system converges; in addition, if two agents belong to the same strongly connected component in the unbounded interaction graph  (called a persistent graph in the present paper), then they converge to the same limit. The convergence rate was provided in \cite{MaGi13} for the system where   the ratio of the reciprocal interaction weights is even allowed to take  a slow diverging value instead of a constant value. In \cite{BoMa15}, a notion of balance condition called balanced asymmetry was proposed, which is stronger than the cut-balance condition, while the balanced asymmetric system includes the cut-balanced system, in which every agent has a positive self-weight, as a special case. The convergence of the system with the balanced asymmetry property is proved under the absolute infinite flow property \cite{ToNe12,ToNe14} for deterministic iterations \cite{BoMa15}.

The arc-balance condition requires that at each time instant the weight of each arc is bounded by a proportional amount of any other arc in the persistent graph. Under this condition, it was proved that the multi-agent system reaches consensus under the condition that the persistent graph contains a directed spanning tree \cite{ShJo13c}. This persistent graph property behaves as forms of network Borel-Cantelli lemmas for consensus algorithms over random graphs \cite{ShAnJo15}. If the persistent graph is strongly connected, the arc balance assumption is a special case of the cut-balance condition imposed on the persistent graph, while in the general case, these two conditions do not cover each other. Note that the results for the discrete-time consensus system under the cut-balance condition in \cite{HeTs13} and the arc-balance condition in \cite{ShJo13c} should be satisfied at each time instant.

In this paper, we   study   discrete-time consensus dynamics  over deterministic networks under relaxed conditions  by allowing that the total amount of the interaction weights  over  each time window of a fixed length satisfies such a condition. Thus the  cut-balance condition is relaxed to the requirement of non-instantaneous reciprocal interactions. We prove that in both cases global consensus is reached if and only if the persistent graph contains a directed spanning tree. In addition, the convergence rate of the system to consensus in both cases are also established in terms of the interactions between agents that have taken place. The technique to prove the result in the cut-balance case is inspired by that used to deal with consensus systems with balanced asymmetry property in \cite{BoMa15} and the cut-balance property with slow divergence of reciprocal weights in \cite{MaGi13}.  It is worth noting that it is only assumed that the self-weight of each agent is bounded by a positive constant from below while the weights between agents can be arbitrary time-varying functions, which relaxes the existing assumptions \cite{ReBe05,CaMoAn08a,NeOlOzTs09,HeTs13}. The critical assumption on the  boundedness of the self-weight of each agent is also discussed and  an illustrative example is provided. Some preliminary results were submitted for presentation at the IEEE Conference on Decision and Control in 2017 \cite{XiShMeCaJo17}.

The rest of the paper is organized as follows. In Section \ref{se:problem}, the global consensus problem is formulated and two main results on the convergence and convergence rates making use of two different balance conditions are given. Section \ref{se:proof1} and Section \ref{se:proof2} present the proofs of the two results, respectively. Section \ref{se:discussion} gives an example to illustrate some critical conditions. The conclusion is drawn in Section \ref{se:conclusion}.

\section{Problem Formulation and Main Results}\label{se:problem}
\subsection{Problem Formulation}
Consider a network with the node set $\mathcal V=\{1,\dots,N\}$, $N\geq2$. Each node $i$ holds a state $x_i(t)\in \mathds{R}$. The initial time is $t_0\geq  0$. The evolution of $x_i(t)$ is given by
\begin{equation}\label{eq:sys1}
{x}_i(t+1)=\sum\limits_{j =1}^N a_{ij}(t)x_j(t),
\end{equation}
where $a_{ij}(t)\geq 0$ stands for the influence of node $j$ on node $i$ at time $t$  and $a_{ii}(t)$ represents the self-confidence of each node. If $a_{ij}(t)>0$ at time $t$, then it is  considered as the weight of arc $(j,i)$ of the graph $\mathbb G(t)=(\mathcal V,\mathcal E(t))$, where $\mathcal E(t)\subseteq\mathcal V\times \mathcal V$.

For the time-varying arc weights $a_{ij}(t)$, we impose the following condition as our standing assumption throughout the paper.

\begin{assumption}\label{ass:1}
For all $i,j\in\mathcal V$ and $t\geq 0$, (i) $a_{ij}(t)\geq 0$; (ii) $\sum_{j=1}^N a_{ij}(t)=1$; (iii) There exists a constant $0<\eta<1$ such that $a_{ii}(t)\geq \eta$.
\end{assumption}

Denote $x(t)=[x_1(t),\dots,x_N(t)]^T$ and $A(t)=[a_{ij}(t)]_{N\times N}$. We know that $A(t)$ is a stochastic matrix from Assumption \ref{ass:1}. System (\ref{eq:sys1}) can be rewritten as
\begin{equation}\label{eq:sys2}
{x}(t+1)=A(t)x(t).
\end{equation}

\begin{remark}
In Assumption \ref{ass:1}, we only assume that the diagonal elements of $A(t)$ are lower bounded by $\eta$, but not requiring  all   nonzero elements of $A(t)$ to be lower bounded by $\eta$, a condition often imposed in the literature \cite{BlHeOlTs05,ReBe05,CaMoAn08a}. It will be seen in later discussions that it will bring many differences and require further efforts for the analysis of the system. The condition that the  diagonal elements of $A(t)$ are lower bounded by $\eta$ is critical for the consensus reaching of system (\ref{eq:sys1}) and its importance will be further illustrated by an example in Section \ref{se:discussion}.
\end{remark}

We continue to introduce the following definition \cite{ShJo13c}.

\begin{definition} An arc $(j,i)$ is called  a persistent  arc   if
\begin{equation}\label{eq:perarc}
\sum_{t=0}^\infty a_{ij}(t)=\infty.
\end{equation}
The set of all persistent arcs is denoted as $\mathcal E_p$ and we call the digraph $\mathbb G_p=(\mathcal V,\mathcal E_p)$ the persistent graph.
\end{definition}

The weight function of each arc in the persistent graph has an infinite $l_1$ norm as can be seen from (\ref{eq:perarc}).
The notions of persistent arcs and persistent graph have also been considered in \cite{HeTs13,MaGi13,MaHe16,BoMa15} for studying the consensus problem of discrete-time and continuous-time systems. In \cite{BoMa15} the persistent graph $\mathbb G_p$ is called an unbounded interactions graph. We will show in the next section that the connectivity of the persistent graph is fundamental for deciding consensus, while those edges whose time-varying interaction weights summing up to a finite number is not critical. The consensus problem considered in this paper is defined as follows.
\begin{definition}
Global consensus is achieved for the considered network if for any initial time $t_0\geq0$, and for any initial value $x(t_0)$, there exists $x_\ast\in \mathds{R}$ such that
$\lim_{t\rightarrow \infty} x_i(t)=x_\ast$ for all $i\in\mathcal V$.
\end{definition}

In addition, we not only derive conditions under which global consensus can be reached, but also characterize the convergence speed in terms of how much interaction  among the nodes has happened in the network.

\subsection{Balance Conditions}

A central aim of this paper is to derive conditions under which the convergence to consensus of system (\ref{eq:sys1}) can be guaranteed by imposing merely the connectivity of the persistent graph. In this case some balance conditions  among the arc weights become essential \cite{ShJo13c,HeTs13}. We introduce the following two balance  conditions.

\begin{assumption} \label{ass:2}
(Balance Condition I) There exist an integer $L\geq 1$ and a constant $K\geq 1$ such that for any $(j,i),(l,k)\in\mathcal E_p$, we have
\begin{equation}\label{eq:ass2}
\sum_{t=s}^{s+L-1}  a_{kl}(t) \leq K  \sum_{t=s}^{s+L-1}  a_{ij}(t)
\end{equation}
for all $s\geq 0$.
\end{assumption}

\begin{assumption} \label{ass:3}
(Balance Condition II) There exist an integer $L\geq 1$ and a constant $K\geq 1$ such that for any nonempty proper subset $S$ of $\mathcal V$, we have
\begin{equation}\label{eq:ass3}
 \sum_{t=s}^{s+L-1}\sum_{i\not\in S,j\in S}  a_{ij}(t) \leq  K\sum_{t=s}^{s+L-1} \sum_{i\in S,j\not\in{S}}  a_{ij}(t)
\end{equation}
for all $s\geq 0$.
\end{assumption}

\begin{remark}
The Balance Condition I is a generalized version of the arc-balance condition introduced in \cite{ShJo13c} where $L=1$. The Balance Condition II is a generalized version of the cut-balance condition introduced in \cite{HeTs13} where $L=1$. These conditions require either the balance between the weights of different persistent arcs or the balance between the amounts of interactions between one group and its remaining part over each time window of a fixed length. When Assumption \ref{ass:2} or  Assumption \ref{ass:3} holds for $L=1$, (\ref{eq:ass2}) or (\ref{eq:ass3}) imposes a restriction on such a balance condition that should be satisfied instantaneously. A relatively large $L$ gives more flexibility on the interaction weights and allows possible non-instantaneous reciprocal interactions between agents.
\end{remark}


\subsection{Main Results}

In this section, we first give some basic observations of the state evolution of system (\ref{eq:sys1}) and then present the main results.

Let
$$
H(t)\doteq\max_{i\in\mathcal V} \{x_i(t)\},\quad h(t)\doteq\min_{i\in\mathcal V}\{x_i(t)\}
$$
be the maximum and minimum state value  at time $t$, respectively. Denote $\Psi\big(t\big)\doteq H(t)-h(t)$ which serves as a metric of consensus.  Note that $\Psi\big(t\big)$ measures the maximum difference among the states of the nodes.

Apparently reaching a consensus of  system (\ref{eq:sys1}) implies that $\lim_{t\rightarrow \infty} \Psi\big(t\big)=0$. In fact the contrary is also true.  It is straightforward to see that $H(t)$ is non-increasing,  $h(t)$ is non-decreasing and thus $\Psi(t)$ is non-increasing. Therefore, for any initial time $t_0\geq0$ and any initial value $x^0=x(t_0)$, there exist $H_\ast, h_\ast\in \mathds{R}$ such that
$$
\lim_{t\rightarrow \infty} H(t)=H_\ast; \quad \lim_{t\rightarrow \infty} h(t)=h_\ast.
$$
If $\lim_{t\rightarrow \infty} \Psi\big(t\big)=0$, we obtain $H_\ast=h_\ast$, which implies that $\lim_{t\rightarrow \infty} x_i(t)=H_\ast$ for all $i\in\mathcal V$.

Let  $\lceil {a} \rceil$ represent the smallest integer that is no less than $a$, and $\lfloor {a} \rfloor$ represent the largest integer that is no greater than $a$.  We present the following two main results, for the two types of balance conditions, respectively.

\begin{theorem}\label{thm:1} Assume that Assumptions \ref{ass:1} and \ref{ass:2} hold.\\
(i) Global consensus is achieved for system (\ref{eq:sys1}) if and only if the persistent graph $\mathbb G_p$ has a directed spanning tree.\\
(ii)  If the persistent graph $\mathbb G_p$ has a directed spanning tree, then for any initial time $t_0\geq0$,  $\epsilon>0$, and $\varepsilon>0$, we have
\begin{align}\label{eq:thm1_1}
\Psi(t)\leq \epsilon \Psi(t_0),\qquad \text{for all } t\geq T_\varepsilon +t^\ast,
\end{align}
where $T_\varepsilon\geq t_0$ such that $\sum_{t=T_\varepsilon}^{\infty}a_{ij}(t) \leq \varepsilon$ for all $(j,i)\in\mathcal E\setminus \mathcal E_p$,
{\small
\begin{equation}\label{eq:thm1_2}
t^\ast\doteq \inf\Bigg\{t\geq 1:  \sum_{k=0}^{t-1} \sum_{j =1, j\neq i, (j,i)\in \mathcal E_p}^N a_{ij}(T_\varepsilon+k)\geq \omega_1d_0(\delta+1)\Bigg\},
\end{equation}
}
$\delta>L(N-1)(1-\eta)$ is a constant,  $d_0$ is the diameter of  $\mathbb G_p$, $\omega_1\doteq\left\lceil\frac{\log{\epsilon}^{-1}}{\log \left(1-\frac{1}{2}{\mathbf Q}^{2d_0}\mathbf R^{d_0}\right)^{-1}}\right\rceil$ with  $\mathbf R \doteq K^{-1}\left[\frac{\delta}{N-1}-L(1-\eta)\right]$, ${\mathbf Q}\doteq e^{- \frac{(N-1)(K(1-\eta+\delta)+L(1-\eta)+\varepsilon)\ln \eta}{\eta-1}}$.
\end{theorem}

\begin{theorem}\label{thm:2} Assume that Assumptions \ref{ass:1} and \ref{ass:3} hold.\\
(i) Global consensus is achieved for system (\ref{eq:sys1}) if and only if the persistent graph $\mathbb G_p$ has a directed spanning tree.\\
(ii) If the persistent graph $\mathbb G_p$ has a directed spanning tree, then for any initial time  $t_0\geq0$ and $\epsilon>0$, we have
\begin{align}\label{eq:thm2_1}
\Psi(t)\leq \epsilon \Psi(t_0),\qquad \text{for all } t\geq k^\ast L+t_0,
\end{align}
where
\begin{align}\label{eq:thm2_2}
&k^\ast\doteq\inf\Bigg\{t\geq 1:\min_{|S(0)|=\cdots=|S(t-1)|}W\sum_{k=0}^{t-1}\sum_{i\not\in S(k+1)\atop j\in S(k)}\sum_{u=0}^{L-1}a_{ij}(kL+u+t_0)\geq \omega_2\left\lfloor\frac{N}{2}\right\rfloor(\eta^L+1)\Bigg\},
\end{align}
with $W=\frac{\eta^L}{(N-1)L}$,
$\omega_2=\left\lceil\frac{\log{\epsilon}^{-1}}{\log \left(1- {{K_\ast}^{-\lfloor\frac{N}{2}\rfloor}}/{{(8N^2)}^{\lfloor\frac{N}{2}\rfloor}}\right)^{-1}}\right\rceil$, $K_\ast=\max\left\{\frac{(N-1)K}{\eta^{L-1}},\frac{N-1}{\eta^L}\right\}$, and $S(k),\ k\geq0,$ being  nonempty proper subsets of $\mathcal V$ with the same cardinality.
\end{theorem}

For both cases, the conclusions (ii) establish the convergence rates of system (\ref{eq:sys1}) to consensus in terms of the interactions between agents having taken place. In the following two sections, we prove these two theorems.

\section{Proof of Theorem \ref{thm:1}}\label{se:proof1}

In this section, we first establish two key technical lemmas, and then present the proofs of Theorem \ref{thm:1}.

\subsection{Key Lemmas}
First we present the following lemma establishing a lower bound for the product of a finite sequence of real numbers.
\begin{lemma}\label{lm:1} Let   $b_k, k=1,\dots,m$ be a sequence of real numbers of length $m$ satisfying $b_k\in [\eta,1]$, $m\geq 0$, where $0<\eta<1$ is a given constant.  Then we have $\prod_{k=1}^m b_k \geq e^{- \frac{\zeta\ln \eta}{\eta-1}}$ if $\sum_{k=1}^m (1-b_k) \leq \zeta$.
\end{lemma}
{\it Proof.} Noticing that $\ln y$ is a concave function on $(0,\infty)$, we obtain
\begin{align*}
\ln y &=\ln \Big[ \frac{y-1}{\eta-1} \cdot \eta +\Big( 1 - \frac{y-1}{\eta-1} \Big)\cdot 1 \Big] \geq \frac{y-1}{\eta-1} \cdot \ln \eta
\end{align*}
for all $y\in [\eta,1]$. Therefore, we conclude that
\begin{align*}
\prod_{k=1}^mb_k&=e^{\sum_{k=1}^m \ln b_k}\geq e^{ {\sum_{k=1}^m \frac{b_k-1}{\eta-1} \cdot \ln \eta}} =e^{-\frac{\ln \eta}{\eta-1} {\sum_{k=1}^m (1-b_k)}} \geq e^{-\frac{\zeta \ln \eta}{\eta-1} }.
\end{align*}
This completes the proof. \hfill$\square$

As will be shown in the following discussions, the fact that the lower bound $e^{- \frac{\zeta\ln \eta}{\eta-1}}$ is independent on $m$ plays a key role in analyzing the node state evolution.

Next, we establish another lemma on the node state evolution.
\begin{lemma}\label{lm:2}
Suppose $x_{i}(s)\leq \mu h(s)+(1-\mu)H(s)$ for some $s\geq t_0$  and $0\leq\mu<1$. Then we have
\begin{align}
x_i(s+\tau)\leq  &\mu\prod_{k=0}^{T-1}a_{ii}(s+k)\cdot h(s)+\Big(1-\mu\prod_{k=0}^{T-1}a_{ii}(s+k)\Big)\cdot H(s)
\end{align}
for all $\tau\leq T$ and $T=0,1,\dots$.
\end{lemma}
{\it Proof.} First we have
\begin{align*}
x_i(s+1)&=\sum\limits_{j =1}^N a_{ij}(s)x_j(s)\nonumber\\
&=a_{ii}(s)x_i(s)+\sum\limits_{j =1, j\neq i}^N a_{ij}(s)x_j(s)\nonumber\\
&\leq a_{ii}(s) \Big[ \mu h(s)+(1-\mu)H(s) \Big]+ \big(1- a_{ii}(s)\big) H(s)\nonumber\\
&= \mu a_{ii}(s) h(s)+\big(1-\mu a_{ii}(s)\big)H(s).
\end{align*}
Recall  that $H(t)$ is non-increasing for all $t$. Thus, iteratively, we obtain
\begin{align*}
& x_i(s+2)\nonumber\\
&=\sum\limits_{j =1}^N a_{ij}(s+1)x_j(s+1)\nonumber\\
&\leq a_{ii}(s+1)x_i(s+1)+\big(1- a_{ii}(s+1)\big)H(s+1)\nonumber\\
&\leq a_{ii}(s+1) \Big[ \mu a_{ii}(s) h(s)+\big(1-\mu a_{ii}(s)\big)H(s)  \Big]+ \big(1- a_{ii}(s+1)\big) H(s)\nonumber\\
&= \mu \prod_{k=0}^{1}a_{ii}(s+k) h(s)+\Big(1-\mu \prod_{k=0}^{1}a_{ii}(s+k)\Big)H(s).
\end{align*}
Moreover, noticing $h(s)\leq H(s)$ and $a_{ii}(t)\in[0,1]$, we also have
\begin{align*}
&x_i(s+1)\nonumber\\
&\leq \mu a_{ii}(s) h(s)+\big(1-\mu a_{ii}(s)\big)H(s)\nonumber\\
&\leq \mu \prod_{k=0}^{1}a_{ii}(s+k) h(s)+\Big(1-\mu \prod_{k=0}^{1}a_{ii}(s+k)\Big)H(s).
\end{align*}

Proceeding the analysis it is straightforward to see  that the desired conclusion holds. \hfill$\square$

\subsection{Proof of Theorem \ref{thm:1} (i)}

\noindent{\it (Sufficiency)}
We introduce
$$
\mathbf{A}_{i}(t)=\sum_{j =1, j\neq i, (j,i)\in \mathcal E_p}^N a_{ij}(t)
$$
for each node $i\in\mathcal V$ and $t\geq0$. According to the definition of the persistent graph, for any initial time $t_0$ and any $\varepsilon>0$, there exists an integer $T_\varepsilon\geq t_0$ such that $\sum_{t=T_\varepsilon}^{\infty}a_{ij}(t) \leq \varepsilon$ for all $(j,i)\in\mathcal E\setminus \mathcal E_p$.

We divide the rest of the proof into four steps.

\vspace{2mm}

\noindent { Step 1.} Take  $T_0=T_\varepsilon$ and  $\delta>L(N-1)(1-\eta)$, where $\eta$ is the constant in Assumption \ref{ass:1} and $L$ is the integer in  Assumption \ref{ass:2}. Let $i_0$ be a root of the persistent graph $\mathbb G_p$ and $(i_0,i_1)\in\mathcal E_p$. Such an $i_1$ exists since $\mathbb G_p$ contains a directed spanning tree. Define
$$
t_1\doteq \inf\big\{t\geq 1:\mbox{$\sum_{k=0}^{t-1} \mathbf{A}_{i_1}(T_0+k) \geq \delta$}\big\}.
$$
Note that $t_1$ is finite since $(i_0,i_1)$ is a persistent arc in $\mathbb G_p$. Let $s$ be the integer satisfying that $(s-1)L\leq t_1<sL$. Since $\delta>L(N-1)(1-\eta)$, one has $s\geq N$. With Assumption \ref{ass:1}, we have that $\sum_{k=0}^{t_1-1} \mathbf{A}_{i_1}(T_0+k) \leq 1-\eta+\delta$.

Since $a_{i_1i_1}(T_0+k)=1- \sum_{j =1, j\neq i_1}^N a_{i_1j}(T_0+k)$ and based on Assumption \ref{ass:1}, we have

(i) $a_{i_1i_1}(T_0+k)\in[\eta,1]$ for all $k=0,\dots,t_1-1$;
  
  {\rm (ii)}
\begin{align*}
 \sum_{k=0}^{t_1-1}\big(1-a_{i_1i_1}(T_0+k) \big)&=\sum_{k=0}^{t_1-1} \mathbf{A}_{i_1}(T_0+k)+\sum_{k=0}^{t_1-1}\sum_{j =1, j\neq i_1,(j,i_1)\not\in\mathcal E_p}^N a_{i_1j}(T_0+k)\\
&\leq  1-\eta+\delta+\varepsilon(N-1).
\end{align*}

Therefore, we conclude from Lemma \ref{lm:1} that
\begin{align}\label{22}
\prod_{k=0}^{t_1-1}a_{i_1i_1}(T_0+k) \geq  e^{- \frac{(1-\eta+\delta+\varepsilon(N-1))\ln \eta}{\eta-1}} \doteq {\mathbf S}.
\end{align}

It is clear from the definition of $\mathbf{A}_{i}(t)$ and the fact $(s-1)L\leq t_1<sL$ that
\begin{align*}
&\sum_{k=0}^{(s-1)L-1} a_{i_1i_r}(T_0+k)\leq\sum_{k=0}^{t_1-1} a_{i_1i_r}(T_0+k)\leq\sum_{k=0}^{t_1-1} \mathbf{A}_{i_1}(T_0+k)\leq 1-\eta+\delta,
\end{align*}
for all $(i_r,i_1)\in\mathcal E_p$.
From Assumption \ref{ass:2}, one has that for any $(j,i)\in\mathcal E_p$,
\begin{align*}
 \sum_{k=0}^{t_1-1} a_{ij}(T_0+k)
&=\sum_{k=0}^{(s-1)L-1} a_{ij}(T_0+k)+\sum_{(s-1)L}^{t_1-1}a_{ij}(T_0+k)\\
&\leq K\sum_{k=0}^{(s-1)L-1} a_{i_1i_r}(T_0+k)+\sum_{(s-1)L}^{sL-1}a_{ij}(T_0+k)\\
&\leq K(1-\eta+\delta)+L(1-\eta).
\end{align*}
For any $i\neq i_1$, it is true that
\begin{equation*}
\begin{split}
&\sum_{k=0}^{t_1-1}\big(1-a_{ii}(T_0+k) \big)\\
&=\sum_{k=0}^{t_1-1} \mathbf{A}_{i}^\ast(T_0+k)+\sum_{k=0}^{t_1-1}\sum_{j =1, j\neq i,(j,i)\not\in\mathcal E_p}^N a_{ij}(t) \\
&\leq  (N-1)(K(1-\eta+\delta)+L(1-\eta)+\varepsilon).
\end{split}
\end{equation*}
Thus in view of Lemma \ref{lm:1}, we have that
\begin{align}\label{eq:thm1_1}
\prod_{k=0}^{t_1-1}a_{ii}(T_0+k) \geq  e^{- \frac{(N-1)(K(1-\eta+\delta)+L(1-\eta)+\varepsilon)\ln \eta}{\eta-1}} = {\mathbf Q},
\end{align}
for $i\neq i_1$. Note that $ {\mathbf Q}< {\mathbf S}$.

Assume that
 \begin{align*}
 x_{i_0}(T_0)\leq \frac{1}{2} h(T_0) +\frac{1}{2}H(T_0).
 \end{align*}
In this step, we establish a bound for $x_{i_0}(T_0+\tau), \tau=0,\dots,t_1$.

Based on Lemma \ref{lm:2}, we obtain
\begin{align}\label{12}
x_{i_0}(T_0+\tau)&\leq  \frac{1}{2}\prod_{k=0}^{t_1-1}a_{i_0i_0}(T_0+k)\cdot h(T_0)+\Big(1-\frac{1}{2}\prod_{k=0}^{t_1-1}a_{i_0i_0}(T_0+k)\Big)\cdot H(T_0)
\end{align}
for all $\tau=0,\dots,t_1$. Then (\ref{eq:thm1_1}) and (\ref{12}) further imply
\begin{align}\label{20}
x_{i_0}(T_0+\tau)\leq  \frac{{{\mathbf Q}}}{2} h(T_0)+\Big(1- \frac{{{\mathbf Q}}}{2} \Big) H(T_0).
\end{align}
for all $\tau=0,\dots,t_1$.

\vspace{2mm}

\noindent {\it Step 2.} In this step, we establish a bound for $x_{i_1}(T_0+t_1)$. Since $\sum_{k=0}^{t_1-1}\mathbf{A}_{i_1}(T_0+k) \geq \delta$, there must exist a node $i_r$ such that $(i_r,i_1)\in\mathcal E_p$ and
\begin{align*}
\sum_{k=0}^{t_1-1}a_{i_1i_r}(T_0+k)  \geq \frac{\delta}{N-1}.
\end{align*}
It follows that
\begin{align*}
\sum_{k=0}^{(s-1)L-1}a_{i_1i_r}(T_0+k)&\geq \frac{\delta}{N-1}-\sum_{(s-1)L}^{t_1-1}a_{i_1i_r}(T_0+k)\nonumber\\
&\geq \frac{\delta}{N-1}-\sum_{(s-1)L}^{sL}a_{i_1i_r}(T_0+k)\\&\geq\frac{\delta}{N-1}-L(1-\eta)\\&>0,\nonumber
\end{align*}
where the last inequality is true since $\delta>L(N-1)(1-\eta)$.
From Assumption \ref{ass:2}, for any arc $(i,j)\in\mathcal E_p$, one has that
\begin{align}\label{eq:thm1_2}
\sum_{k=0}^{t_1-1} a_{ij}(T_0+k) &\geq\sum_{k=0}^{(s-1)L-1} a_{ij}(T_0+k)\nonumber\\
&\geq K^{-1} \sum_{k=0}^{(s-1)L-1}a_{i_1i_r}(T_0+k)\nonumber\\ & \geq K^{-1}\left[\frac{\delta}{N-1}-L(1-\eta)\right]\nonumber\\
&= \mathbf R,
\end{align}
where $\mathbf R$ is defined in Theorem \ref{thm:1}. The above inequality also holds for the arc $(i_0,i_1)$ since $(i_0,i_1)\in\mathcal E_p$.

First according to (\ref{20}), we have
\begin{align*}
x_{i_1}(T_0+1)&=\sum\limits_{j =1}^N a_{i_1j}(T_0)x_j(T_0)\nonumber\\
&\leq a_{i_1i_0}(T_0)x_{i_0}(T_0)+ \big(1- a_{i_1i_0}(T_0)\big) H(T_0)\nonumber\\
&\leq a_{i_1i_0}(T_0) \Big[  \frac{{{\mathbf Q}}}{2} h(T_0)+\Big(1- \frac{{{\mathbf Q}}}{2} \Big) H(T_0) \Big]+  \big(1- a_{i_1i_0}(T_0)\big) H(T_0)\nonumber\\
&= \frac{{{\mathbf Q}}}{2}a_{i_1i_0}(T_0) h(T_0)+\Big(1- \frac{{{\mathbf Q}}}{2}a_{i_1i_0}(T_0) \Big) H(T_0).
\end{align*}
Then  for $T_0+2$, we have
{\small
\begin{align*}
&x_{i_1}(T_0+2)\nonumber\\
&=\sum\limits_{j =1}^N a_{i_1j}(T_0+1)x_j(T_0+1)\nonumber\\
&\leq a_{i_1i_0}(T_0+1)x_{i_0}(T_0+1)+ a_{i_1i_1}(T_0+1)x_{i_1}(T_0+1) +\big(1- a_{i_1i_0}(T_0+1)-a_{i_1i_1}(T_0+1)\big) H(T_0+1)\nonumber\\
&\leq a_{i_1i_0}(T_0+1) \Big[  \frac{{{\mathbf Q}}}{2} h(T_0)+\Big(1- \frac{{{\mathbf Q}}}{2} \Big) H(T_0) \Big]\nonumber\\
   & \ \ +a_{i_1i_1}(T_0+1)\bigg[\frac{{{\mathbf Q}}}{2}a_{i_1i_0}(T_0) h(T_0)+\Big(1- \frac{{{\mathbf Q}}}{2}a_{i_1i_0}(T_0) \Big) H(T_0)\bigg ]\nonumber\\
   &\ \  +\big(1- a_{i_1i_0}(T_0+1)-a_{i_1i_1}(T_0+1)\big) H(T_0)\nonumber\\
&= \frac{{{\mathbf Q}}}{2}\Big[a_{i_1i_0}(T_0+1)+a_{i_1i_1}(T_0+1)a_{i_1i_0}(T_0)\Big] h(T_0) +\bigg[1- \frac{{{\mathbf Q}}}{2}\Big[a_{i_1i_0}(T_0+1)+a_{i_1i_0}(T_0+1)a_{i_1i_0}(T_0)\Big] \bigg]H(T_0).
\end{align*}
}
By induction it is straightforward to find that
{\small
\begin{align}\label{eq:thm1_6}
&x_{i_1}(T_0+t_1)\nonumber\\
&\leq \frac{{{\mathbf Q}}}{2}\Big[\sum_{\tau=0}^{t_1-1} \prod_{k=\tau+1}^{t_1-1} a_{i_1i_1}(T_0+k)a_{i_1i_0}(T_0+\tau)\Big]h(T_0) +\bigg[1- \frac{{{\mathbf Q}}}{2}\Big[\sum_{\tau=0}^{t_1-1} \prod_{k=\tau+1}^{t_1-1} a_{i_1i_1}(T_0+k)a_{i_1i_0}(T_0+\tau)\Big] \bigg]H(T_0)\nonumber\\
&\leq \frac{{{\mathbf Q}}}{2}\Big(\prod_{k=0}^{t_1-1} a_{i_1i_1}(T_0+k)\Big) \Big( \sum_{k=0}^{t_1-1} a_{i_1i_0}(T_0+k)\Big) h(T_0)+\bigg[1- \frac{{{\mathbf Q}}}{2}\Big(\prod_{k=0}^{t_1-1} a_{i_1i_1}(T_0+k)\Big) \Big( \sum_{k=0}^{t_1-1} a_{i_1i_0}(T_0+k)\Big)\bigg]H(T_0)\nonumber\\
&\leq \frac{1}{2}{\mathbf S}{\mathbf Q}\mathbf R  h(T_0)+ \Big( 1-\frac{1}{2}{\mathbf S}{\mathbf Q}\mathbf R \Big)H(T_0),
\end{align}
}
where the last inequality is due to (\ref{22}) and (\ref{eq:thm1_2}).

\vspace{2mm}

\noindent {\it Step 3.} Let $\mathcal V_0=\{i_0\}$ and $\mathcal V_1=\{i:\ (i_0,i)\in\mathcal E_p\}$. It is obvious from (\ref{eq:thm1_6}) and in view of (\ref{eq:thm1_1}) that for any $i\in\mathcal V_1$,
\begin{align*}
x_{i}(T_0+t_1)\leq \frac{1}{2}{\mathbf Q}^2\mathbf R  h(T_0)+ \Big( 1-\frac{1}{2}{\mathbf Q}^2\mathbf R \Big)H(T_0),
\end{align*}
Let $\mathcal V_2$ be a subset of $\mathcal V\backslash(\mathcal V_0\cup\mathcal V_1)$ and consist of all the nodes each of which has a neighbor in $\mathcal V_0\cup\mathcal V_1$ in $\mathbb G_p$. We continue to define
$$t_2\doteq \inf\big\{t\geq t_1+1:  \sum_{k=t_1}^{t-1} \mathbf{A}_{i_1}(T_0+k) \geq \delta\big\}.$$
Similarly, one can find an integer $s$ such that $(s-1)L\leq t_2-t_1<sL.$ In this step, we will give an upper bound for $x_i(T_0+t_2)$ for $i\in\mathcal V_0\cup\mathcal{V}_1\cup\mathcal V_2.$

Similar to the calculations of (\ref{22}) and  (\ref{eq:thm1_1}) in  step 1, one can derive that
\begin{align}\label{eq:thm1_3}
\prod_{k=t_1}^{t_2-1}a_{i_1i_1}(T_0+k) \geq {\mathbf S}.
\end{align}
and
\begin{align}\label{eq:thm1_4}
\prod_{k=t_1}^{t_2-1}a_{ii}(T_0+k) \geq  {\mathbf Q},\ \  i\neq i_1.
\end{align}
Using Lemma \ref{lm:2} and noting that ${\mathbf S}>{\mathbf Q}$,  we obtain
\begin{align*}
x_{i}(T_0+t_1+\tau)\leq \frac{1}{2}{\mathbf Q}^3\mathbf R  h(T_0)+ \Big( 1-\frac{1}{2}{\mathbf Q}^3\mathbf R \Big)H(T_0),
\end{align*}
for $i=i_0,i_1,$ and $\tau=0,\ldots,t_2-t_1$.

For any $i_2\in\mathcal V_2,$ there is an arc $(i,i_2)\in\mathcal E_p$ for some $i\in\mathcal V_0\cup\mathcal V_1$. Similar to (\ref{eq:thm1_2}), Assumption \ref{ass:2} implies that
\begin{align}\label{eq:thm1_5}
\sum_{k=t_1}^{t_2-1} a_{i_2i}(T_0+k)  \geq K^{-1} \sum_{k=t_1}^{t_1+(s-1)L-1}a_{i_1i_r}(T_0+k)  \geq \mathbf R.
\end{align}
for some $(i_r,i_1)\in\mathcal E_p$. Following similar calculations of $x_{i_1}(T_0+t_1)$ in step 2, we obtain
{\small
\begin{align}
&x_{i_2}(T_0+t_2)\nonumber\\
&\leq \frac{1}{2}{\mathbf Q}^3\mathbf R \Big(\prod_{k=t_1}^{t_2-1} a_{i_2i_2}(T_0+k)\Big) \Big( \sum_{k=t_1}^{t_2-1} a_{i_2i}(T_0+k)\Big) h(T_0)\nonumber\\
&\ \ \ +\bigg[1- \frac{1}{2}{\mathbf Q}^3\mathbf R\Big(\prod_{k=t_1}^{t_2-1} a_{i_2i_2}(T_0+k)\Big) \Big( \sum_{k=t_1}^{t_2-1} a_{i_2i}(T_0+k)\Big)\bigg]H(T_0)\nonumber\\
&\leq \frac{1}{2}{\mathbf Q}^4\mathbf R^2 h(T_0)+ \Big( 1-\frac{1}{2}{\mathbf Q}^4\mathbf R^2 \Big)H(T_0).
\end{align}
}

\noindent {\it Step 4.} Continuing this process, $\mathcal V_3,\ldots,\mathcal V_{d_0}$ can be defined similarly with $d_0$ being the diameter of $\mathbb G_p$ and a time sequence $t_1,\ldots,t_{d_0}$ can be defined as
$${t}_r\doteq \inf\Big\{t\geq t_{r-1}+1:  \sum_{k=t_{r-1}}^{t-1} \mathbf{A}_{i_1}(T_0+k) \geq \delta\Big\},$$
for $r=1,2,\ldots,d_0,$ with $t_0=0$. It is easy to see that the root $i_0$ can be selected such that $\cup_{i=0}^{d_0}\mathcal V_i=\mathcal V$. The bound for $x_i(T_0+t_{d_0})$ can be established as
\begin{align}
x_{i}(T_0+t_{d_0})\leq \frac{1}{2}{\mathbf Q}^{2d_0}\mathbf R^{d_0}  h(T_0)+ \Big( 1-\frac{1}{2}{\mathbf Q}^{2d_0}\mathbf R^{d_0}\Big)H(T_0),
\end{align}
for all $i=1,\ldots,N$.
A bound for $\Psi(T_0+t_{d_0})$ is thus derived
$$\Psi(T_0+t_{d_0})\leq\Big( 1-\frac{1}{2}{\mathbf Q}^{2d_0}\mathbf R^{d_0} \Big)\Psi(T_0).$$
When $x_{i_0}(T_0)> \frac{1}{2} h(T_0) +\frac{1}{2}H(T_0)$, one can establish a lower bound for $x_{i}(T_0+t_{d_0})$ by a symmetric argument and derive the same inequality for $\Psi(T_0+t_{d_0})$ as above.

\vskip 2mm
Repeating the above estimate, one can find an infinite increasing time sequence $t_1,\ldots,t_{d_0},t_{d_0+1},\ldots,t_{2d_0},\ldots,$ defined by
\begin{equation}\label{eq:tr}
{t}_r\doteq \inf\big\{t\geq t_{r-1}+1:  \sum_{k=t_{r-1}}^{t-1} \mathbf{A}_{i_1}(T_0+k) \geq \delta\big\},
\end{equation}
and we have
\begin{align}\label{eq:100}
\Psi(T_0+t_{rd_0})\leq\Big( 1-\frac{1}{2}{\mathbf Q}^{2d_0}\mathbf R^{d_0}\Big)^r\Psi(T_0),
\end{align}
for $r=1,2,\ldots.$ It implies that the sequence $\Psi(T_0+t_{rd_0}),\ r=1,2,\ldots,$ converges to 0 as $r$ goes to infinity. Since $\Psi(T_0+t_{rd_0})$ is a subsequence of a non-increasing sequence $\Psi(t),\ t\geq0$, $\Psi(t)$ converges to 0 as $t$ goes to infinity as well, which completes the proof.

{\it (Necessity)} The proof of the necessity part is similar to that of Theorem 3.1 in \cite{ShJo13c} and is thus omitted here.


\subsection{Proof of Theorem \ref{thm:1} (ii)}

 Note that from the definition of $t_r$ in (\ref{eq:tr}) and the definition of $\mathbf{A}_{i_1}$, one knows that for any $r\geq1$,
$$\sum_{k=t_{r-1}}^{t_r-1} \mathbf{A}_{i_1}(T_\varepsilon+k)\leq 1+\delta.$$
It follows that
$$\sum_{k=0}^{t_{\omega_1}-1} \mathbf{A}_{i_1}(T_\varepsilon+k)\leq \omega_1d_0(1+\delta).$$
By the definition of $t^\ast$ in (\ref{eq:thm1_2}), $t^\ast\geq t_{\omega_1}d_0$. For $t\geq T_{\varepsilon}+t^\ast$, applying (\ref{eq:100}) we have
\begin{align*}
\Psi(t)&\leq\Psi(T_{\varepsilon}+t^\ast)\leq \Psi(T_{\varepsilon}+t_{\omega_1}d_0)\\
&\leq \Big(1-\frac{1}{2}{\mathbf Q}^{2d_0}\mathbf R^{d_0}\Big)^{\omega_1}\Psi(T_{\varepsilon})\\
&\leq \epsilon \Psi(T_{\varepsilon}).
\end{align*}
\hfill $\Box$

\section{Proof of Theorem \ref{thm:2}}\label{se:proof2}
In this section, we first establish some technical preliminaries, and then establish the convergence statement  Theorem \ref{thm:2} (i)    and   the contraction rate of $\Psi(t)$  claimed in Theorem \ref{thm:2} (ii).

\subsection{Technical Preliminaries}
Consider system (\ref{eq:sys1}) with the initial time $t_0$. Let $y(t)=x(tL+t_0)$ and $B(t)=A((t+1)L-1+t_0)\cdots A(tL+1+t_0)A(tL+t_0)$. Then the dynamics of $y$-system is given by
\begin{equation}\label{eq:y}
y(t+1)=B(t)y(t).
\end{equation}
Letting $\Phi(t)\doteq\max_{i\in\mathcal V} y_i(t)-\min_{i\in\mathcal V} y_i(t)$, one has that $\Phi(t)=\Psi(tL+t_0)$. One can conclude that $\lim_{t\rightarrow\infty}\Psi(t)=0$ if and only if $\lim_{t\rightarrow\infty}\Phi(t)=0$ since $\Psi(t)$ is a nonincreasing function of $t$. Hence we  establish the global consensus of system (\ref{eq:sys1}) by studying the property of the $y$-system (\ref{eq:y}).

We first establish  two technical  lemmas, whose proofs are given in Appendices  \ref{ap:A} and  \ref{ap:B}, respectively.

\begin{lemma}\label{lm:4}
Let $A_1,A_2,\ldots,A_m$ be stochastic matrices and for each $A_i,\ 1\leq i\leq m$, assume that all the diagonal elements are no less than $\eta,\ 0<\eta<1$.  Let $B_m=A_1A_2\cdots A_m$ and $C_m=A_1+\cdots+A_m$.  Then we have
\begin{equation}\label{eq:lm4}
\sum_{i\in S,j\not\in S}  (B_m)_{ij}\geq \eta^{m-1}\sum_{i\in S,j\not\in S}  (C_m)_{ij},
\end{equation}
where $S$ is an arbitrary nonempty proper subset of $\mathcal V$ and $(B_m)_{ij}$ is the $ij$-th element of $B_m$.
\end{lemma}

\begin{lemma}\label{lm:5}
Let $A_1,A_2,\ldots,A_m$ be $N\times N$ stochastic matrices, $B_m=A_1A_2\cdots A_m$ and $C_m=A_1+\cdots+A_m$. Then we have
\begin{equation}\label{eq:lm5}
\sum_{i\in S,j\not\in S}  (B_m)_{ij}\leq (N-1)\sum_{i\in S,j\not\in S}  (C_m)_{ij},
\end{equation}
where $S$ is an arbitrary nonempty proper subset of $\mathcal V$.
\end{lemma}

We derive some useful properties of the system matrix $B(t)$ in (\ref{eq:y}) based on Assumption \ref{ass:3} in the following lemma.

\begin{lemma}\label{lm:3}
If Assumptions \ref{ass:1} and \ref{ass:3} hold, then each matrix $B(t),\ t\geq0$, has positive diagonals lower bounded by $\eta^L$ and satisfies the cut-balance condition
\begin{equation}\label{eq:CT1}
\sum_{i\not\in S,j\in S}b_{ij}(t)\leq M_\ast\sum_{i\in S,j\not\in S}b_{ij}(t)
\end{equation}
for any nonempty proper subset $S$ of $\mathcal V$ with $M_\ast=(N-1)K\eta^{-L+1}$. Let $\mathbb G_p^\prime=(\mathcal V,\mathcal E_p^\prime)$ be a directed graph where $(j,i)\in\mathcal E_p^\prime$ if and only if $\sum_{t=0}^\infty b_{ij}(t)=\infty$. The persistent graph $\mathbb G_p$ contains a directed spanning tree if and only if $\mathbb G_p^\prime$ contains a directed spanning tree.
\end{lemma}
{\it Proof.} Since $a_{ii}(t)\geq\eta$ for all $i\in\mathcal V, t\geq0$, it is obvious that $b_{ii}(t)\geq\eta^L$ for all $t\geq 0$. Applying Lemmas \ref{lm:4} and \ref{lm:5} to system matrices  $A(t)$ and $B(t)$ in (\ref{eq:sys1}) and (\ref{eq:y}) and in view of Assumption \ref{ass:3}, one has
\begin{align*}
\sum_{i\not\in S,j\in S}  b_{ij}(t)&\leq (N-1)\sum_{i\not\in S,j\in S}\sum_{u=0}^{L-1}a_{ij}(tL+u+t_0)\\
&\leq(N-1)K\sum_{i\in S,j\not\in S}\sum_{u=0}^{L-1}a_{ij}(tL+u+t_0)\\
&\leq(N-1)K\eta^{-L+1}\sum_{i\in S,j\not\in S}  b_{ij}(t).
\end{align*}
Hence (\ref{eq:CT1}) holds.

(Sufficiency) Suppose that $(j,i)$ is an arc of $\mathbb G_p$. By the definition of a persistent arc, $\sum_{t=0}^\infty a_{ij}(t)=\infty$. There must exist a time sequence $t_{k_1}, t_{k_2},\dots,$ diverging to infinity with nonnegative integers $k_1<k_2<\cdots$ such that $k_sL\leq t_{k_s}\leq (k_s+1)L-1,\ s\geq1$ and $\sum_{s=1}^\infty a_{ij}(t_{k_s}+t_0)=\infty$. One has that
\begin{align*}
b_{ij}(k_s)&\geq a_{ii}((k_s+1)L-1)\cdots a_{ij}(t_{k_s}+t_0)\cdots a_{jj}(k_sL+t_0)\\
&\geq \eta^{L-1}a_{ij}(t_{k_s}+t_0),\ s\geq1.
\end{align*}
It follows that $\sum_{t=0}^\infty b_{ij}(t)\geq \eta^{L-1}\sum_{s=1}^\infty a_{ij}(t_{k_s}+t_0)=\infty$, implying that $(j,i)$ is a persistent arc of $\mathbb G_p^\prime$. $\mathbb G_p^\prime$  contains a directed spanning tree since $\mathbb G_p$ does.

(Necessity) Suppose that $(j,i)$ is an arc of $\mathbb G_p^\prime$. Note that $\sum_{t=0}^\infty b_{ij}(t)=\infty$ and
\begin{align*}
b_{ij}(t)=&\sum_{k_1,\ldots,k_{L-1}\in\mathcal V}a_{ik_{L-1}}((t+1)L-1)\\
&\cdots a_{k_2k_1}(tL+1+t_0)a_{k_1j}(tL+t_0).
\end{align*}
It follows that there exist integers $k_1,\ldots,k_{L-1}\in\mathcal V$ such that $\sum_{t=0}^\infty a_{ik_{L-1}}((t+1)L-1+t_0)\cdots a_{k_2k_1}(tL+1+t_0)a_{k_1j}(tL+t_0)=\infty$. Since $a_{ij}(t)\leq1$ for all $i,j\in\mathcal V,t\geq0$, one has that $\sum_{t=0}^\infty a_{k_{s+1}k_s}(tL+s+t_0)=\infty$ for all $0\leq s\leq L-1$ with $k_0=j$ and $k_L=i$, implying that $(k_s,k_{s+1})\in\mathcal E_p$. This implies that there exists a directed path from node $j$ to $i$ in $\mathbb G_p$. Hence if $\mathbb G_p^\prime$ contains a directed spanning tree, so does $\mathbb G_p$.
\hfill $\Box$

\begin{remark}
It has been proved in \cite{HeTs13} that under the cut-balance condition (\ref{eq:CT1}) if $\mathbb G_p^\prime$ contains a directed spanning tree then it is strongly connected. Following a similar argument, one can show that when Assumption \ref{ass:3} holds, if $\mathbb G_p$ contains a directed spanning tree then it is strongly connected.
\end{remark}

Consider the system
\begin{equation}\label{eq:y2}
y(t+1)=B(t)y(t),
\end{equation}
where $B(t)=[b_{ij}(t)]\in \mathds{R}^{N\times N},\ b_{ij}(t)\geq0,$ and $\sum_{j=1}^Nb_{ij}(t)=1$. The following lemma is a convergence result of the cut-balanced system.

\begin{lemma}\cite{MaHe16}\label{lm:MaHe16}
For system (\ref{eq:y2}), suppose that the following assumptions hold:
\begin{itemize}
\item There exists a $\gamma>0$ such that $b_{ii}(t)\geq\gamma$ for all $i\in\mathcal V,t\geq0$.
\item There exists a constant $M_\ast$ such that for every $t$ and nonempty proper subset $S$ of $\mathcal V$, there holds
\begin{equation}\label{eq:CT2}
\sum_{i\not\in S,j\in S}b_{ij}(t)\leq M_\ast\sum_{i\in S,j\not\in S}b_{ij}(t).
\end{equation}
\end{itemize}
Then $\lim_{t\rightarrow\infty}y_i(t)$ exists for every $i$. Let $\mathbb G_p^\prime=(\mathcal V,\mathcal E_p^\prime)$ be the persistent graph where $(j,i)\in\mathcal E_p^\prime$ if $\sum_{k=0}^\infty b_{ij}(t)=\infty$. If $\mathbb G_p^\prime$ contains a directed spanning tree, then global consensus is reached.
\end{lemma}

Lemma \ref{lm:MaHe16} is a special case of Theorem 1 in \cite{ToLa14} restricted to deterministic systems. The result has also been proved in Theorem 2 in \cite{BoMa15} for balanced asymmetric systems which include the system in Lemma \ref{lm:MaHe16} as a special case and we will introduce in the next subsection. In Lemma \ref{lm:MaHe16}, the condition that $\mathbb G_p^\prime$ contains a directed spanning tree is also necessary for the global consensus of system (\ref{eq:y2}), which has been proved in Theorem 2 in \cite{BoMa15}.

\subsection{Proof of Theorem \ref{thm:2} (i)}\label{sec:proof2_1}
 Lemma \ref{lm:3} shows that the $y$-system (\ref{eq:y}) satisfies the assumptions of Lemma \ref{lm:MaHe16}. One concludes that $\mathbb G_p^\prime$ defined in Lemma \ref{lm:3} contains a directed spanning tree if and only if global consensus of system (\ref{eq:y}) is reached. Combining with Lemma \ref{lm:3}, the conclusion of Theorem \ref{thm:2}. (i) immediately follows.


\subsection{Proof of Theorem \ref{thm:2} (ii)}

In this subsection, we provide a contraction rate of $\Phi(t)$ and hence a corresponding contraction rate of $\Psi(t)$ can be obtained. We have seen that system (\ref{eq:y}) satisfies the cut-balanced condition (\ref{eq:CT2}). Instead of considering the cut-balanced system, we consider a system with $B(t)$ satisfying the balanced asymmetric condition. As will be seen shortly, the balanced asymmetric condition includes the cut-balanced condition with $b_{ii}(t)$ lower bounded by a positive constant as a special case.

\begin{assumption}\label{ass:4} (Balanced Asymmetry) \cite{BoMa15} There exists a constant $M\geq1$ such that for any two nonempty proper subsets $S_1,S_2$ of $\mathcal V$ with the same cardinality, the matrices $B(t),\ t\geq0,$ satisfy that
\begin{equation}\label{eq:BA}
\sum_{i\not\in S_1,j\in S_2}b_{ij}(t)\leq M\sum_{i\in S_1,j\not\in S_2}b_{ij}(t).
\end{equation}
\end{assumption}

\begin{remark}
As pointed out in Remark 1 in \cite{BoMa15}, the balanced asymmetry condition is stronger than the cut-balance condition (\ref{eq:CT2}). But if $B(t)$ has positive diagonal elements lower bounded by a positive constant $\gamma$ and satisfies (\ref{eq:CT2}), then it satisfies the balanced asymmetry condition with $M=\max\{M_\ast,\frac{N-1}{\gamma}\}.$
\end{remark}

In the following, we consider system (\ref{eq:y2}) and assume that the matrices $B(t),\ t\geq0,$ satisfy the balanced asymmetry condition. We first establish the convergence rate of $\Phi(t)=\max_{i\in\mathcal V}y_i(t)-\min_{i\in\mathcal V}y_i(t)$ and then apply the result to the cut-balanced system (\ref{eq:y}). We introduce the notion of absolute infinite flow property \cite{ToNe12,BoMa15} which has a close relationship with the connectivity of persistent graphs.

\begin{definition}
The sequence of matrices $B(t),\ t\geq0$ is said to have the absolute infinite flow property if the following holds
\begin{equation}\label{eq:AIF}
\sum_{t=0}^\infty\Big(\sum_{i\not\in S(t+1)\atop j\in S(t)}b_{ij}(t)+\sum_{i\in S(t+1)\atop j\not\in S(t)}b_{ij}(t)\Big)=\infty
\end{equation}
for every sequence $S(t),\ t\geq0,$ of nonempty proper subsets of $\mathcal V$ with the same cardinality.
\end{definition}

If the matrix sequence $B(t),\ t\geq0,$ has the absolute infinite flow property and satisfies the balanced asymmetry condition, we can define an infinite time sequence $t_0,t_1,t_2,\dots$ based on (\ref{eq:AIF}). Let $t_0^0=t_0$ and define a finite time sequence $t_p^0,t_p^1,\dots,t_p^{\lfloor\frac{N}{2}\rfloor},\ p\geq0$. $t_p^{q+1}$ is defined by
\begin{align}\label{eq:condition2}
t_p^{q+1}\doteq&\inf\Bigg\{t\geq t_p^q+1:\nonumber\\
&\min_{|S(t_p^q)|=\cdots=|S(t-1)|}\sum_{k=t_p^q}^{t-1}\sum_{i\not\in S(k+1)\atop j\in S(k)}b_{ij}(k)\geq1\Bigg\},
\end{align}
where $|S|$ denotes the cardinality of a set $S$. Let $t_{p+1}=t_p^{\lfloor\frac{N}{2}\rfloor}$ and $t_{p+1}^0=t_{p+1}$. We derive an infinite time sequence $t_0,t_1,t_2,\dots$. Since (\ref{eq:BA}) holds, one has that for every sequence $S(t),\ t\geq0,$ of nonempty proper subsets of $\mathcal V$ with the same cardinality
$$\sum_{t=0}^\infty\sum_{i\not\in S(t+1)\atop j\in S(t)}b_{ij}(t)=\infty,$$
from which it is clear that (\ref{eq:condition2}) is well-defined.

\begin{proposition}\label{lm:BA}
For system (\ref{eq:y2}), assume that the sequence of matrices $B(t),t\geq0,$ satisfies Assumption \ref{ass:4}. If it has the absolute infinite flow property, then
\begin{equation}\label{eq:Brate}
\Phi(t_{p+1})\leq \Big(1- {{M}^{-\lfloor\frac{N}{2}\rfloor}}/{{(8N^2)}^{\lfloor\frac{N}{2}\rfloor}}\Big)\Phi(t_p).
\end{equation}
and global consensus of system (\ref{eq:y2}) is reached.
\end{proposition}

We introduce some new notations and lemmas for the proof of Proposition \ref{lm:BA}. For $t\geq0$, let $\sigma_t$ be a permutation of $\mathcal V$ such that for $i<j$, either $y_{\sigma_t(i)}(t)<y_{\sigma_t(j)}(t)$ or $y_{\sigma_t(i)}(t)=y_{\sigma_t(j)}(t)$ and $\sigma_t(i)<\sigma_t(j)$ holds. Define $z_i(t)\doteq y_{\sigma_t(i)}(t),\ t\geq0$. From the definition of the permutation $\sigma_t$, one knows that for all $t\geq0,$ if $i<j,$  then $z_i(t)\leq z_j(t)$. Hence $z(t)=[z_1(t),\dots,z_N(t)]^T$ is a sorted state vector.

\begin{remark}
Inequality (\ref{eq:Brate}) for the contraction rate of $\Psi(t)$ takes the same form as that in Proposition 2 in \cite{MaGi13} which deals with a continuous-time system under persistent connectivity.
We will employ similar ideas to derive (\ref{eq:Brate}). The dynamics of the continuous-time system considered in \cite{HeTs13,MaGi13} is
$$\dot{y}_i(t)=\sum_{j=1}^Nb_{ij}(t)(y_j(t)-y_i(t)).$$
The solution to the system is a locally absolutely continuous function $y$ that satisfies the integral equation
$$y_i(t)=y_i(0)+\int_0^t\sum_{j=1}^Nb_{ij}(s)(y_j(s)-y_i(s))ds.$$
A key property of the continuous-time system proved in \cite{HeTs13b} is that the sorted state $z(t)$ satisfies an equation of the same form as the state $y(t)$
$$z_i(t)=z_i(0)+\int_0^t\sum_{j=1}^Nc^\prime_{ij}(s)(z_j(s)-z_i(s))ds,$$
where $c^\prime_{ij}(t)\doteq b_{\sigma_t(i),\sigma_t(j)}(t)$. In addition, if $B(t),\ t\geq0,$ satisfy the cut-balance condition, then $C^\prime(t)=[c^\prime_{ij}(t)]_{N\times N},\ t\geq0,$ satisfy the cut-balance condition as well \cite{MaGi13}. However, for the discrete-time system (\ref{eq:y2}), $z(t)$ does not satisfy the equation of the same form as $y(t)$ with the above notations. We  modify the definition of $c^\prime_{ij}(t)$ such that this still holds.
\end{remark}

Define $c_{ij}(t)\doteq b_{\sigma_{t+1}(i),\sigma_t(j)}(t)$. It is obvious that $\sum_{j=1}^Nc_{ij}(t)=1$ for all $i\in\mathcal V$, $t\geq0$. In view of the definition $z_i(t)=y_{\sigma_t(i)}(t)$ for $t\geq0$, one has
\begin{align}\label{eq:z}
z_i(t+1)&=y_{\sigma_{t+1}(i)}(t+1)=\sum_{j=1}^nb_{\sigma_{t+1}(i),j}(t)y_j(t)\nonumber\\
&=\sum_{j=1}^nb_{\sigma_{t+1}(i),\sigma_t(j)}(t)y_{\sigma_t(j)}(t)
=\sum_{j=1}^Nc_{ij}(t)z_j(t).
\end{align}
In addition, the interaction weights $c_{ij}(t)$ have the following property.

\begin{lemma}\label{lm:8}
Assume that $B(t),t\geq 0$ satisfy Assumption \ref{ass:4}. For any nonempty proper subsets $S_1,\ S_2$ of $\mathcal V$ with the same cardinality, $c_{ij}(t)$ satisfies
\begin{equation}\label{eq:c}
\sum_{i\not\in S_1,j\in S_2}c_{ij}(t)\leq M\sum_{i\in S_1,j\not\in S_2}c_{ij}(t).
\end{equation}
\end{lemma}
{\it Proof.} For a fixed $t$, let $S_3=\{\sigma_{t+1}(i): i\in S_1\}$ and $S_4=\{\sigma_t(j): j\in S_2\}$. Since $\sigma_t,\sigma_{t+1}$ are permutations of $\mathcal V$, $S_3$ and $S_4$ have the same cardinality and using (\ref{eq:BA}), one has
\begin{align*}
\sum_{i\not\in S_1,j\in S_2}c_{ij}(t)&=\sum_{i\not\in S_1,j\in S_2}b_{\sigma_{t+1}(i),\sigma_t(j)}(t)\nonumber\\
&=\sum_{i\not\in S_3,j\in S_4}b_{ij}(t)\leq M\sum_{i\in S_3,j\not\in S_4}b_{ij}(t)\nonumber\\
&=M\sum_{i\in S_1,j\not\in S_2}c_{ij}(t).
\end{align*}\hfill $\Box$

\begin{remark}
Note that if $B(t),\ t\geq0,$ satisfy the cut-balance condition, $C(t)=[c_{ij}(t)]_{N\times N},\ t\geq0,$ do not preserve the cut-balance property in general while $C^\prime(t),\ t\geq0,$ do. However, the evolution of $z_i(t)$ does not satisfy $z_i(t+1)=\sum_{j=1}^Nc^\prime_{ij}(t)z_j(t),\ i\in\mathcal V,$ in general. In this case, though $C^\prime(t)=[c^\prime_{ij}(t)]_{N\times N},\ t\geq0,$ satisfy the cut-balance condition, the evolution of $z_i(t)$ cannot be directly expressed using $C^\prime(t)$ and is not easy to be analyzed making use of the property of $C^\prime(t)$. In addition, the diagonal elements of the newly defined matrix $C(t),\ t\geq0$, are not necessarily positive any more and some existing results in the literature cannot be directly applied to the system (\ref{eq:z}).
\end{remark}

\begin{example}
Consider the system (\ref{eq:y2}) consisting of four agents.  Let
\begin{align*}
&B(t)=\begin{cases}\begin{bmatrix}
         \frac{1}{2} & 0 & \frac{1}{2} & 0 \\
         0 & \frac{1}{2} & 0 & \frac{1}{2} \\
         \frac{1}{2} & 0 & \frac{1}{2} & 0 \\
         0 & \frac{1}{2} & 0 & \frac{1}{2} \\
       \end{bmatrix},&\text{if } t \text{ is even},\\
        \begin{bmatrix}
         \frac{1}{2} & \frac{1}{2} & 0 & 0 \\
         \frac{1}{2} & \frac{1}{2} & 0 & 0 \\
         0 & 0 & \frac{1}{2} & \frac{1}{2} \\
         0 & 0& \frac{1}{2} & \frac{1}{2} \\
                \end{bmatrix},&\text{if } t \text{ is odd}.
                \end{cases}
                \end{align*}
 Let the initial state of the system be $y(0)=[0,1,2,3]^T$. We only consider the first step evolution of the system as shown in Fig.~\ref{fig:11}.

 \begin{figure} [htbp]
\begin{center}
\includegraphics[width=8.5cm]{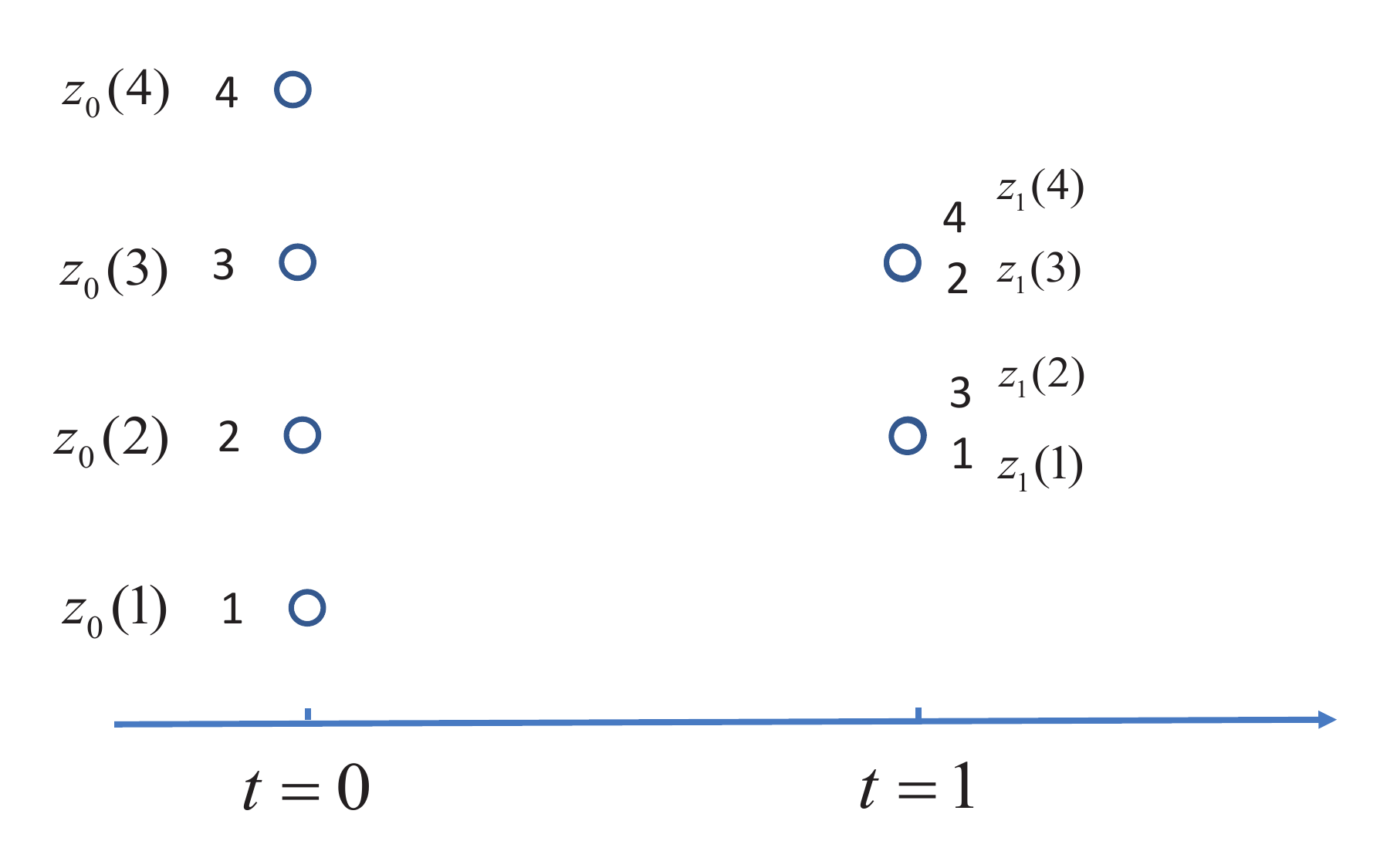}    
\caption{The system state evolution at the first step.}\label{fig:11}
\end{center}
\end{figure}

One can easily see that
$$y(1)=B(0)y(0)=[1,2,1,2]^T.$$
Since $y(0)$ is already sorted, $\sigma_0(i)=i,\ i=1,\dots,4$ and $z(0)=y(0)$. By the definition of $\sigma_t$, we have
$\sigma_1(1)=1,\sigma_1(2)=3, \sigma_1(3)=2,\sigma_1(4)=4,\ \text{and } z(1)=[1,1,2,2]^T.$
By the definitions of $c^\prime_{ij}(t)\doteq b_{\sigma_t(i),\sigma_t(j)}(t)$ and $c_{ij}(t)\doteq b_{\sigma_{t+1}(i),\sigma_t(j)}(t)$, one has
$$C^\prime(0)= B(0),$$
and
\begin{align*}
C(0)&= [b_{\sigma_{1}(i),\sigma_0(j)}(0)]_{N\times N}\\
&=
\begin{bmatrix}
         b_{11} & b_{12} & b_{13} & b_{14} \\
         b_{31} & b_{32} & b_{33} & b_{34} \\
         b_{21} & b_{22} & b_{23} & b_{24} \\
         b_{41} & b_{42} & b_{43} & b_{44} \\
       \end{bmatrix}=\begin{bmatrix}
         \frac{1}{2} & 0 & \frac{1}{2} & 0 \\
         \frac{1}{2} & 0 & \frac{1}{2} & 0 \\
         0 & \frac{1}{2} & 0 & \frac{1}{2} \\
         0 & \frac{1}{2} & 0 & \frac{1}{2} \\
       \end{bmatrix}.
       \end{align*}
 Note that the matrix $B(0)$ has positive diagonal elements while $C(0)$ does not.     It can be directly verified that $z(1)\neq y(1)=C^\prime(0)z(0)$ and $z(1)=C(0)z(0)$.
       \hfill $\Box$
\end{example}

\begin{lemma}\label{lm:9}
Assume that the matrix $B$ satisfies that
\begin{equation}\label{eq:BA2}
\sum_{i\not\in S_1,j\in S_2}b_{ij}\leq M\sum_{i\in S_1,j\not\in S_2}b_{ij}(t),
\end{equation}
for a constant $M\geq1$ and any two nonempty proper subsets $S_1,S_2$ of $\mathcal V$ with the same cardinality. Let $\sigma$ and $\mu$ be permutations of $\mathcal V$ and $c_{ij}=b_{\mu(i),\sigma(j)}$.  Then for any sorted vector $z\in \mathds R^n$ and $1\leq l\leq N-1$, one has
\begin{equation}\label{eq:lm9-1}
\sum_{i=1}^lM^{-i}\Big(\sum_{j=1}^Nc_{ij}(z_j-z_i)\Big)\geq
(z_{l+1}-z_l)M^{-l}\sum_{i=l+1}^N\sum_{j=1}^lc_{ji}\geq0.
\end{equation}
\end{lemma}

\begin{remark}
The proof of Lemma \ref{lm:9} is similar to that of Lemma 2  in \cite{HeTs13} and Lemma 9 in \cite{MaGi13} and hence is omitted here.  Note that  if the matrix $B$ only satisfies the cut-balance condition (\ref{eq:BA}), then the inequality (\ref{eq:lm9-1}) may not hold since the  matrix $C=[c_{ij}]_{N\times N}$ defined in Lemma \ref{lm:9} does not satisfy the cut-balance condition any more in general.
\end{remark}

{\it Proof of Proposition \ref{lm:BA}.} Note that $z_i(t)$ satisfies $z_i(t+1)=\sum_{j=1}^Nc_{ij}(t)z_j(t),\ i\in\mathcal V$. In addition, $z(t)=[z_1(t),\dots,z_N(t)]^T$ is a sorted state vector and $\Phi(t)=z_n(t)-z_1(t)$ for $t\geq0$. With the key inequality (\ref{eq:lm9-1}) in Lemma \ref{lm:9} in hand, using similar ideas to the proofs of Lemmas 10, 11, and Proposition 2 in Section 4.2 in \cite{MaGi13}, one can derive (\ref{eq:Brate}). \hfill $\Box$

Next consider system  (\ref{eq:y2}) with $B(t)$ satisfying the cut-balance condition (\ref{eq:CT2}) and $b_{ii}(t)\geq\gamma$ for all $i\in\mathcal V,t\geq0$. We show that when the persistent graph $\mathbb G_p^\prime$ contains a directed spanning tree, then the matrix sequence  $B(t),\ t\geq0,$ has the absolute infinite flow property.  First note that under the cut-balance condition, if the persistent graph contains a directed spanning tree then it is strongly connected. For every sequence $S(t),\ t\geq0,$ of nonempty proper subsets of $\mathcal V$, if there are an infinite number of pairs of  $S(t)$ and $S(t+1)$ such that $S(t)\neq S(t+1)$, then for each of this pair, one has
$$\sum_{i\not\in S(t+1)\atop j\in S(t)}b_{ij}(t)\geq\gamma,$$
since $b_{ii}(t)\geq\gamma$ for all $i\in\mathcal V$, $t\geq0$. It follows that
$$\sum_{t=0}^{\infty}\sum_{i\not\in S(t+1)\atop j\in S(t)}b_{ij}(t)=\infty.$$
If there are only a finite number of pairs of  $S(t)$ and $S(t+1)$  such that $S(t)\neq S(t+1)$, then there exists an integer $T_0$ such that for $t\geq T_0$, $S(t)=S$. It follows that
$$\sum_{t=0}^{\infty}\sum_{i\not\in S(t+1)\atop j\in S(t)}b_{ij}(t)=\sum_{t=T_0}^{\infty}\sum_{i\not\in S\atop j\in S}b_{ij}(t).$$
Since the persistent graph $\mathbb G_p^\prime$ is strongly connected, there must exist an arc from $S$ to $\bar{S}$ and one concludes that the above expression is equal to $\infty$. One concludes that the matrix sequence $B(t),\ t\geq0,$ has the absolute infinite flow property. Then we can define a time sequence $t_0,t_1,\dots$ based on (\ref{eq:condition2}) for the cut-balanced system in the same way as for the balanced asymmetric system. Note that  when $B(t),\ t\geq0,$ satisfy the cut-balance condition (\ref{eq:CT2}),  they also satisfy the balanced asymmetry condition with $M=\max\{M_\ast,\frac{N-1}{\gamma}\}$. We immediately have the following proposition by applying Proposition \ref{lm:BA}.

\begin{proposition}\label{co:11}
For system (\ref{eq:y2}), assume that the matrices $B(t),\ t\geq0,$ satisfy the cut-balance condition (\ref{eq:CT2}) and $b_{ii}(t)\geq\gamma$ for all $i\in\mathcal V,t\geq0$. If the persistent graph $\mathbb G_p^\prime$ contains a directed spanning tree, then
\begin{equation}\label{eq:co11}
\Phi(t_{p+1})\leq \Big(1-{{M}^{-\lfloor\frac{N}{2}\rfloor}}/{{(8N^2)}^{\lfloor\frac{N}{2}\rfloor}}\Big)\Phi(t_p),
\end{equation}
where $M=\max\{M_\ast,\frac{N-1}{\gamma}\}$ and global consensus is reached.
\end{proposition}

\begin{remark}
Proposition \ref{co:11} gives a convergence rate of the system (\ref{eq:y2}) satisfying the two assumptions in Lemma \ref{lm:MaHe16}. Note that the proof of the convergence result for the consensus system under non-instantaneous reciprocal interactions in \cite{MaHe16} made use of the intermediate result Lemma \ref{lm:MaHe16}. With the help of Proposition \ref{co:11}, one can relate the convergence rate of the system discussed in  \cite{MaHe16} to the amount of interactions having taken place as well.
\end{remark}

{\it Proof of Theorem \ref{thm:2} (ii):}  For system (\ref{eq:sys1}) and any given initial time $t_0\geq0$, let $k_0^0=k_0=0$ and define a finite time sequence $k_p^0,k_p^1,\dots,k_p^{\lfloor\frac{N}{2}\rfloor},\ p\geq0$. $k_p^{q+1}$ is defined by
\begin{align}\label{eq:condition1}
&k_p^{q+1}\doteq\inf\Big\{t\geq k_p^q+1:\nonumber\\
&\min_{|S(k)|=\cdots=|S(t-1)|}W\sum_{k=k_p^q}^{t-1}\sum_{i\not\in S(k+1)\atop j\in S(k)}\sum_{u=0}^{L-1}a_{ij}(kL+u+t_0)\geq1\Big\},
\end{align}
where $W=\frac{\eta^L}{(N-1)L}$ is a constant. Let $k_{p+1}=k_p^{\lfloor\frac{N}{2}\rfloor}$ and $k_{p+1}^0=k_{p+1}$. We derive an infinite time sequence $k_0,k_1,k_2,\dots$. Under Assumptions \ref{ass:1} and \ref{ass:3}, it can be shown that when the persistent graph $\mathbb G_p$ contains a directed spanning tree, the time sequence $k_0,k_1,k_2,\dots$ is well-defined.

We first show that if the persistent graph $\mathbb G_p$ contains a directed spanning tree, then
\begin{equation}\label{eq:prop12}
\Psi(k_{p+1}L+t_0)\leq \Big(1-{{K_\ast}^{-\lfloor\frac{N}{2}\rfloor}}/{{(8N^2)}^{\lfloor\frac{N}{2}\rfloor}}\Big)\Psi(k_pL+t_0),
\end{equation}
where $K_\ast=\max\{\frac{(N-1)K}{\eta^{L-1}},\frac{N-1}{\eta^L}\}$.

Consider system (\ref{eq:y}) derived based on system (\ref{eq:sys1}). It has been shown in Lemma \ref{lm:3} that system (\ref{eq:y}) satisfies the assumptions of Proposition \ref{co:11} with $M_\ast=(N-1)K\eta^{-L+1}$ and $\gamma=\eta^L$. Next we verify that $k_p^{q+1}$ defined in (\ref{eq:condition1}) satisfies that
$$\sum_{k=k_p^q}^{k_p^{q+1}-1}\sum_{i\not\in S(k+1)\atop j\in S(k)}b_{ij}(k)\geq1.$$

Let $S_1,S_2$ be two nonempty proper subsets of $\mathcal V$ with the same cardinality. If $S_1=S_2$, then it follows from Lemma \ref{lm:4} that
$$\eta^{L-1}\sum_{i\not\in S_1\atop j\in S_1}\sum_{u=0}^{L-1}a_{ij}(kL+u+t_0)\leq \sum_{i\not\in S_1\atop j\in S_1}b_{ij}(k).$$
If $S_1\neq S_2$, then $\sum_{i\not\in S_1\atop j\in S_2}\sum_{u=0}^{L-1}a_{ij}(kL+u+t_0)\leq (N-1)L$ and $\sum_{i\not\in S_1\atop j\in S_2}b_{ij}(k)\geq\eta^L$ since $b_{ii}(k)\geq\eta^L$ for all $i\in\mathcal V,\ k\geq0$. This implies that
$$\sum_{i\not\in S_1\atop j\in S_2}\sum_{u=0}^{L-1}a_{ij}(kL+u+t_0)\leq (N-1)L\leq \frac{(N-1)L}{\eta^L}\sum_{i\not\in S_1\atop j\in S_2}b_{ij}(k).$$
Hence for all $S_1,S_2$ and $k\geq0$, it always holds that
\begin{align*}
& \frac{\eta^L}{(N-1)L}\sum_{i\not\in S_1\atop j\in S_2}\sum_{u=0}^{L-1}a_{ij}(kL+u+t_0)\\
&=W\sum_{i\not\in S_1\atop j\in S_2}\sum_{u=0}^{L-1}a_{ij}(kL+u+t_0)\leq \sum_{i\not\in S_1\atop j\in S_2}b_{ij}(k).
\end{align*}
Combining with (\ref{eq:condition1}), one has that
\begin{align*}
&\sum_{k=k_p^q}^{k_p^{q+1}-1}\sum_{i\not\in S(k+1)\atop j\in S(k)}b_{ij}(k)\\
&\geq W\sum_{k=k_p^q}^{k_p^{q+1}-1}\sum_{i\not\in S(k+1)\atop j\in S(k)}\sum_{u=0}^{L-1}a_{ij}(kL+u+t_0)\geq1.
\end{align*}

Note that $\Phi(t)=\Psi(tL+t_0)$ and applying (\ref{eq:co11}) in Proposition \ref{co:11} immediately gives (\ref{eq:prop12}).

Next we prove (\ref{eq:thm2_1}).  Note that for any $k\geq0$ and any sequence $S(k),\ k\geq0,$ of nonempty proper subsets of $\mathcal V$ with the same cardinality, it always holds that
$$W\sum_{i\not\in S(k+1)\atop j\in S(k)}\sum_{u=0}^{L-1}a_{ij}(kL+u+t_0)\leq WL(N-1)=\eta^L.$$
It follows  from the definition of $k_p^{q+1}$ in (\ref{eq:condition1}) that for any sequence $S(k),\ k\geq0,$ of nonempty proper subsets of $\mathcal V$ with the same cardinality, and any $p\geq0,\ 0\leq q\leq \left\lfloor\frac{N}{2}\right\rfloor-1$,
$$W\sum_{k=k_p^{q}}^{k_p^{q+1}-1}\sum_{i\not\in S(k+1)\atop j\in S(k)}\sum_{u=0}^{L-1}a_{ij}(kL+u+t_0)\leq \eta^L+1.$$
Therefore,
$$W\sum_{k=0}^{k_{\omega_2}-1}\sum_{i\not\in S(k+1)\atop j\in S(k)}\sum_{u=0}^{L-1}a_{ij}(kL+u+t_0)\leq \omega_2\left\lfloor\frac{N}{2}\right\rfloor(\eta^L+1).$$
By the definition of (\ref{eq:thm2_2}), $k^\ast\geq k_{\omega_2}$. Applying (\ref{eq:prop12}), one has that if $t\geq k^\ast L+t_0$, then
\begin{align*}
\Psi(t)&\leq\Psi(k^\ast L+t_0)\leq \Psi(k_{\omega_2} L+t_0)\\
&\leq \Big(1-\frac{{K_\ast}^{-\lfloor\frac{N}{2}\rfloor}}{{(8N^2)}^{\lfloor\frac{N}{2}\rfloor}}\Big)^{\omega_2}\Psi(t_0)\\
&\leq \epsilon \Psi(t_0).
\end{align*}
This proves the desired contraction rate.
\hfill $\Box$



\section{Discussions}\label{se:discussion}

For the consensus system (\ref{eq:sys1}), the assumption that the nonzero elements of $A(t)$ are lower bounded by a positive constant $\eta$ is often imposed \cite{BlHeOlTs05,ReBe05,CaMoAn08a}.  Assumption \ref{ass:1} has relaxed this by discarding the requirement on the positive lower boundness for the off-diagonal elements of $A(t)$. However, the existence of $\eta$ as a lower bound for $a_{ii}(t)$ is critical for the convergence to consensus of system (\ref{eq:sys1}). Next we give an example to illustrate that if the diagonal elements are not lower bounded by $\eta$, then consensus may not be reached under the same conditions as in Theorem \ref{thm:2}.

\begin{example} Consider a three-agent system. Assume that the interaction graph switches periodically among three graphs $\mathbb G_1,\ \mathbb G_2,$ and $\mathbb G_3$ given in Fig.~\ref{fig:1_1}. Let the initial time $t_0=1.$  The system matrix $A(t)$ is given by
\begin{align*}
&A(3k+1)=\begin{bmatrix}
  \frac{1}{3k+1} & 1-\frac{1}{3k+1} & 0 \\
  1-\frac{1}{(3k+1)^2} & \frac{1}{(3k+1)^2} & 0 \\
  0 & 0 & 1 \\
\end{bmatrix},\\
&A(3k+2)=\begin{bmatrix}
  \frac{1}{(3k+2)^2} & 0 & 1-\frac{1}{(3k+2)^2}  \\
  0  & 1 & 0\\
  1-\frac{1}{3k+2} & 0  & \frac{1}{3k+2}
  \end{bmatrix},\\
  & A(3k+3)=\begin{bmatrix}
    1  & 0 & 0 \\
  0  &  \frac{1}{3k+3} & 1-\frac{1}{3k+3}  \\
  0 &  1-\frac{1}{(3k+3)^2} & \frac{1}{(3k+3)^2} \\
\end{bmatrix},
\end{align*}
for $k\geq0$. Note that though the matrix $A(t)$ has positive diagonals for all $t\geq1$, there does not exist a positive constant $\eta>0$ such that $a_{ii}(t)\geq\eta$ for all $t\geq1$ since $A(3k+r)$ has some positive element converging to 0 for all $r=1,2,3$ as $k\rightarrow\infty$.

\begin{figure} [htbp]
\begin{center}
\includegraphics[width=9cm]{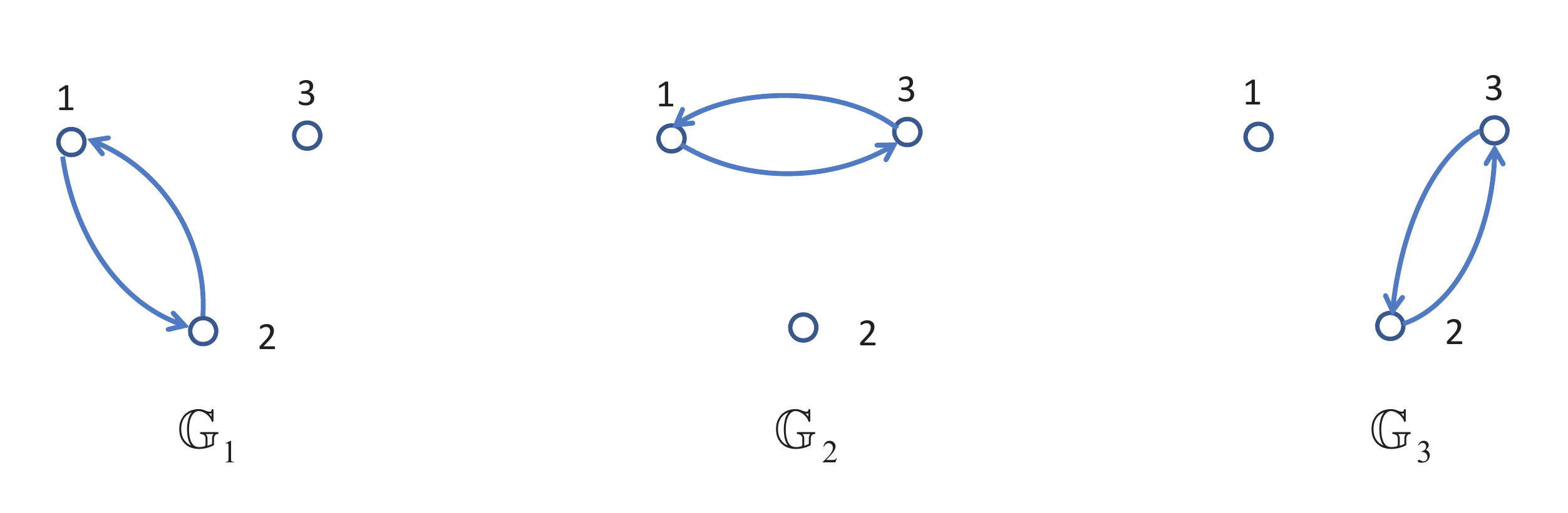}    
\caption{The interaction graph switches periodically among $\mathbb G_1,\mathbb G_2,$ and $\mathbb G_3$.}\label{fig:1_1}
\end{center}
\end{figure}

One can verify that the matrix sequence $A(t),\ t\geq1,$ satisfies Assumption \ref{ass:3} with  $K=2$ and $L=1$  in (\ref{eq:ass3}) since $\frac{1-\frac{1}{t^2}}{1-\frac{1}{t}}=\frac{t+1}{t}\leq 2$ for all $t\geq1$. However, it does not satisfy the balanced asymmetry condition in (\ref{eq:BA}). To see this, consider the matrix $A(3k+1),\ k\geq0,$ and let $S_1=\{1\}$ and $S_2=\{2\}$. It is easy to see that
$$\sum_{i\not\in S_2\atop j\in S_1}a_{ij}(3k+1)=\frac{1}{3k+1}=(3k+1)\sum_{i\in S_2\atop j\not\in S_1}a_{ij}(3k+1).$$
One concludes that the sequence $A(t),\ t\geq1$ does not satisfy the balanced asymmetry condition since $3k+1$ is not bounded as $k\rightarrow\infty.$

 It is obvious that the persistent graph $\mathbb G_p$ is strongly connected. In addition, the matrix sequence $A(t),\ t\geq1,$ has the absolute infinite flow property. To verify this, one only has to consider the sequence $S(t),\ t\geq1,$ of sets with the cardinality equal to 1 since $\mathcal V\backslash S(t)$ also appears in the definition of absolute infinite flow property and there are 3 agents in total. Assume that each $S(t)$ has the cardinality equal to 1. For any $t=3k+1,\ k\geq0$ and the set $S(3k+1)=\{1\}$, one can see that
\begin{equation}\label{eq:ex1}
\sum_{i\not\in S(3k+2)\atop j\in S(3k+1)}a_{ij}(3k+1)+\sum_{i\in S(3k+2)\atop j\not\in S(3k+1)}a_{ij}(3k+1)\geq\frac{1}{3k+1},
\end{equation}
for any $S(3k+2)$. For $S(3k+1)=\{2\}$, the above inequality also holds for any $S(3k+2)$. For $S(3k+1)=\{3\}$, if $S(3k+2)$ is $\{1\}$ or $\{2\}$, then the left hand side of (\ref{eq:ex1}) is at least 2; if $S(3k+2)=\{3\}$, then the left hand side of (\ref{eq:ex1}) is 0, in which case it is clear that for any $S(3k+3)$,
$$\sum_{i\not\in S(3k+3)\atop j\in S(3k+2)}a_{ij}(3k+2)+\sum_{i\in S(3k+3)\atop j\not\in S(3k+2)}a_{ij}(3k+2)\geq\frac{1}{3k+2}.$$
To sum up, in all cases one has
$$\sum_{r=1}^3\Big(\sum_{i\not\in S(3k+r+1)\atop j\in S(3k+r)}a_{ij}(3k+r)+\sum_{i\in S(3k+r+1)\atop j\not\in S(3k+r)}a_{ij}(3k+r)\Big)\geq\frac{1}{3k+2},$$
for all nonempty proper subset $S(3k+r)$ of $\mathcal V$ satisfying $|S(3k+r)|=|S(3k+r+1)|,\ r=1,2,3,$ and $k\geq0$, which implies that
$$\sum_{t=1}^\infty\Big(\sum_{i\not\in S(t+1)\atop j\in S(t)}a_{ij}(t)+\sum_{i\in S(t+1)\atop j\not\in S(t)}a_{ij}(t)\Big)\geq\sum_{k=0}^\infty\frac{1}{3k+2}=\infty,$$
for all nonempty proper sequence $S(t),\ t\geq1,$ of subsets of $\mathcal V$ with the same cardinality. One concludes that the matrix sequence $A(t),\ t\geq1,$ has the absolute infinite flow property.

\begin{figure} [htbp]
\begin{center}
\includegraphics[width=8cm]{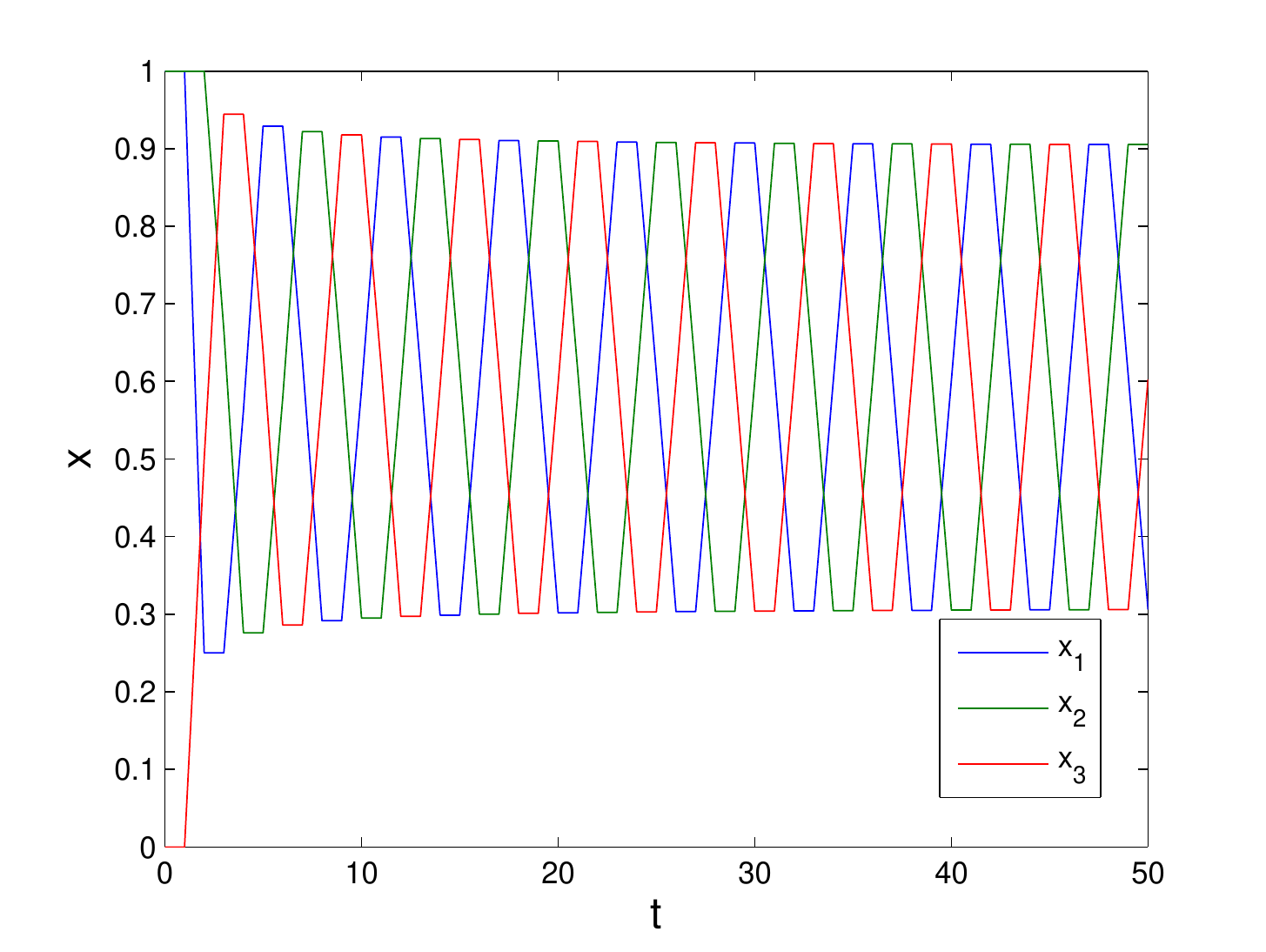}    
\caption{The system state does not reach a consensus.}\label{fig:1_2}
\end{center}
\end{figure}

We next show that global consensus cannot be reached. Consider the initial condition $x(1)=[1, 1,0]^T$. It is obvious that $\Psi(1)=\Psi(2)=1$. For $t\geq2$, one can show that
$$\Psi(t+1)\geq(1-\frac{1}{t^2}) \Psi(t).$$
It follows that
$$\lim_{t\rightarrow\infty}\Psi(t)\geq\prod_{t=2}^\infty(1-\frac{1}{t^2})M(2)=\prod_{t=2}^\infty(1-\frac{1}{t^2})>0,$$
since $\sum_{t=2}^\infty\frac{1}{t^2}<\infty$. The evolution of the system state is depicted in Fig.~\ref{fig:1_2}, which illustrates the disagreement of the system states.
\hfill $\Box$
\end{example}
\section{Conclusions}\label{se:conclusion}

In this paper, we have generalized the cut-balance and arc-balance conditions in the literature so as to allow for non-instantaneous reciprocal interactions between agents. The assumption on the existence of a lower bound on the nonzero weights $a_{ij}$ of the arcs has been relaxed. Illustrative examples have been provided to show the necessity of imposing a positive lower bound on the self-weights of the agents. It has been shown that global consensus is reached if and only if the persistent graph contains a directed spanning tree. The estimate of the convergence rate of the discrete-time system has been given which is not established for the cut-balance case in \cite{MaHe16}. Future work may consider multi-agent systems consisting of agents interacting with each other through attractive and repulsive couplings \cite{Al13,ShJoJo13,He14,XiCaJo16}.

\medskip

\medskip

\section*{Appendix}

\subsection*{A. Proof of Lemma \ref{lm:4}}\label{ap:A}
 We show by induction that
\begin{equation}\label{eq:lmcut1_1}
\sum_{i\in S,j\in\bar{S}}  (B_l)_{ij}\geq \eta^{l-1}\sum_{i\in S,j\in\bar{S}}  (C_l)_{ij},
\end{equation}
for $1\leq l\leq m$. For the matrix $B_2=A_1A_2$, one has
\begin{align}\label{eq:lmcut1_2}
\sum_{i\in S,j\in\bar{S}}  (B_2)_{ij}&=\sum_{i\in S,j\in\bar{S}}\sum_{k=1}^N(A_1)_{ik}(A_2)_{kj}\nonumber\\
&=\sum_{k\in S}\sum_{i\in S,j\in\bar{S}}(A_1)_{ik}(A_2)_{kj}+\sum_{k\in \bar{S}}\sum_{i\in S,j\in\bar{S}}(A_1)_{ik}(A_2)_{kj}\nonumber\\
&\geq\sum_{k\in S}\sum_{j\in\bar{S}}(A_1)_{ii}(A_2)_{kj}+\sum_{k\in \bar{S}}\sum_{i\in S}(A_1)_{ik}(A_2)_{jj}\nonumber\\
&\geq\eta\sum_{k\in S,j\in\bar{S}}(A_2)_{kj}+\eta\sum_{k\in \bar{S},i\in S}(A_1)_{ik}\nonumber\\
&=\eta\sum_{i\in S,j\in\bar{S}}(C_2)_{ij}
\end{align}
Thus (\ref{eq:lmcut1_1}) holds for $l=2$. If $m=2$, then the proof is complete.

Suppose that $m>2$. Assume that (\ref{eq:lmcut1_1}) is true for $l\in\{2,\ldots,s\}$, where $s\in\{2,\ldots,m-1\}$. Since the diagonal elements of $A_i$ are at least $\eta$, one has $(B_{s})_{ii}\geq\eta^{s}$ for $1\leq i\leq N$. Noting that $B_{s+1}=B_sA_{s+1}$ and $C_{s+1}=C_{s+1}+A_{s+1}$, we have
\begin{align}
\sum_{i\in S,j\in\bar{S}}(B_{s+1})_{ij}&=\sum_{k\in S}\sum_{i\in S,j\in\bar{S}}(B_s)_{ik}(A_{s+1})_{kj}+\sum_{k\in \bar{S}}\sum_{i\in S,j\in\bar{S}}(B_s)_{ik}(A_{s+1})_{kj}\nonumber\\
&\geq\eta^s\sum_{k\in S,j\in\bar{S}}(A_{s+1})_{kj}+\eta\sum_{k\in \bar{S},i\in S}(B_s)_{ik}\nonumber\\
&\geq\eta^{s}\sum_{k\in S,j\in\bar{S}}(A_{s+1})_{kj}+\eta^s\sum_{k\in \bar{S},i\in S}(C_s)_{ik}\nonumber\\
&=\eta^{s}\sum_{i\in S,j\in\bar{S}}(C_{s+1})_{ij}.
\end{align}
Hence, (\ref{eq:lmcut1_1}) holds for $l=s+1$. Therefore, (\ref{eq:lmcut1_1}) holds for $1\leq l\leq m$ by induction.\hfill $\Box$

\subsection*{B. Proof of Lemma \ref{lm:5}}\label{ap:B}
It will be shown by induction that
\begin{equation}\label{eq:lmcut2_1}
\sum_{i\in S,j\in\bar{S}}  (B_l)_{ij}\leq (N-1)\sum_{i\in S,j\in\bar{S}}  (C_l)_{ij},
\end{equation}
for $1\leq l \leq m$. In view of (\ref{eq:lmcut1_2}) and the fact that $A_1$ and $A_2$ are stochastic matrices, one has
\begin{align*}
\sum_{i\in S,j\in\bar{S}}  (B_2)_{ij}&=
\sum_{k\in S}\sum_{i\in S,j\in\bar{S}}(A_1)_{ik}(A_2)_{kj}+\sum_{k\in \bar{S}}\sum_{i\in S,j\in\bar{S}}(A_1)_{ik}(A_2)_{kj}\\
&\leq|S|\sum_{k\in S}\sum_{j\in\bar{S}}(A_2)_{kj}+\sum_{k\in \bar{S}}\sum_{i\in S}(A_1)_{ik}\\
&\leq(N-1)\sum_{i\in S,j\in \bar{S}}(A_2)_{ij}+\sum_{i\in S,j\in \bar{S}}(A_1)_{ij}\\
&\leq(N-1)\sum_{i\in S,j\in \bar{S}}(C_2)_{ij}.
\end{align*}
which implies that (\ref{eq:lmcut2_1}) holds for $l=2.$

Suppose that $m>2$. Assume that (\ref{eq:lmcut2_1}) holds for $l\in\{2,\ldots,s\}$, where $s\in\{2,\ldots,m-1\}$. Noting that $B_{s+1}=B_sA_{s+1}$ and $B_s$ is a stochastic matrix, one has
\begin{align}
&\sum_{i\in S,j\in\bar{S}}(B_{s+1})_{ij}\nonumber\\
&=\sum_{k\in S}\sum_{i\in S,j\in\bar{S}}(B_s)_{ik}(A_{s+1})_{kj}+\sum_{k\in \bar{S}}\sum_{i\in S,j\in\bar{S}}(B_s)_{ik}(A_{s+1})_{kj}\nonumber\\
&\leq(N-1)\sum_{k\in S,j\in\bar{S}}(A_{s+1})_{kj}+\sum_{k\in \bar{S},i\in S}(B_s)_{ik}\nonumber\\
&\leq(N-1)\sum_{k\in S,j\in\bar{S}}(A_{s+1})_{kj}+(N-1)\sum_{k\in \bar{S},i\in S}(C_s)_{ik}\nonumber\\
&\leq\eta^{s}\sum_{i\in S,j\in\bar{S}}(C_{s+1})_{ij}.
\end{align}
Hence, (\ref{eq:lmcut1_1}) holds for $l=s+1$. Therefore, (\ref{eq:lmcut1_1}) holds for $1\leq l\leq m$ by induction.\hfill $\Box$


\end{document}